\documentclass[onecolumn]{emulateapj}

\newcommand{\Lsun}{{\hbox {L$_\odot$}}}
\newcommand{\Msun}{{\hbox {M$_\odot$}}}

\def\13co{$^{13}$CO}
\def\c18o{C$^{18}$O}

\slugcomment{Accepted in the Astrophysical Journal , 2005,
January, 2}

\shorttitle{Far-infrared spectroscopy of NGC~1068}
\shortauthors{Spinoglio et al.}

\begin{document}


\title{The far-infrared emission line and continuum spectrum of the Seyfert
galaxy NGC~1068\thanks{ISO is an ESA project with instruments
funded by ESA Member States (especially the PI countries: France,
Germany, the Netherlands and the United Kingdom) and with the
participation of ISAS and NASA.}}
\author{Luigi Spinoglio}\affil{Istituto di Fisica dello Spazio
Interplanetario, CNR, via Fosso del Cavaliere 100, I-00133 Roma,
Italy} \email{luigi@ifsi.rm.cnr.it}

\author{Matthew A. Malkan}\affil{Physics \& Astronomy Dept.,
UCLA, Los Angeles, CA 90095, USA} \email{malkan@astro.ucla.edu}

\author{Howard A. Smith} \affil{Harvard-Smithsonian CfA, 60 Garden St.,
Cambridge, MA 02138, USA} \email{hsmith@cfa.harvard.edu}

\author{Eduardo Gonz\'alez-Alfonso} \affil{Universidad de Alcal\'a de Henares,
Departamento de F\'{\i}sica, Campus Universitario, E-28871 Alcal\'a de Henares,
Madrid, Spain} \email{eduardo.gonzalez@uah.es}

\author{Jacqueline Fischer} \affil{Naval Research Laboratory, Code 7213,
Washington DC 20375, USA} \email{Jackie.Fischer@nrl.navy.mil}

\begin{abstract}
We report on the analysis of the first complete far-infrared
spectrum (43-197$\mu$m) of the Seyfert 2 galaxy NGC~1068 as
observed with the {\it Long Wavelength Spectrometer} (LWS)
onboard the {\it Infrared Space Observatory} (ISO).  In addition
to the 7 expected ionic fine structure emission lines, the OH
rotational lines at 79, 119 and 163$\mu$m were all detected in
emission, which is unique among galaxies with full LWS spectra,
where the 119$\mu$m line, when detected, is always in absorption.
The observed line intensities were modelled together with ISO
{\it Short Wavelength Spectrometer} (SWS) and optical and
ultraviolet line intensities from the literature, considering two
independent emission components: the AGN component and the starburst
component in the circumnuclear ring of $\sim$ 3 kpc in size. Using
the UV to mid-IR emission line spectrum to constrain the nuclear
ionizing continuum, we have confirmed previous results: a
canonical power-law ionizing spectrum is a poorer fit than one
with a deep absorption trough, while the presence of a {\it big
blue bump} is ruled out.  Based on the instantaneous starburst age
of 5 Myr constrained by the Br $\gamma$ equivalent width in the
starburst ring, and starburst synthesis models of the mid- and
far-infrared fine-structure line emission, a low ionization
parameter (U=10$^{-3.5}$) and low densities (n=100 cm$^{-3}$) are
derived. Combining the AGN and starburst components, we succeeded
in modeling the overall UV to far-IR atomic spectrum of NGC~1068,
reproducing the line fluxes to within a factor 2.0 on average
with a standard deviation of 1.3,
and the overall continuum as the sum of the contribution of
the thermal dust emission in the ionized and neutral components.
The OH 119 $\mu$m
emission indicates that the line is collisionally excited, and
arises in a warm and dense region. The OH emission has been
modeled using spherically symmetric, non-local, non-LTE radiative
transfer models. The models indicate that the bulk of the
emission arises from the nuclear region, although some extended
contribution from the starburst is not ruled out. The OH
abundance in the nuclear region is expected to be $\sim10^{-5}$,
characteristic of X-ray dominated regions.
\end{abstract}


\section{INTRODUCTION}

NGC~1068 is known as the archetypical Seyfert type~2 galaxy.  It
is nearby, luminous \citep[${\rm L_{IR}=2 \times 10^{11}
L_{\odot}}$][]{bla97}, and it has been extensively observed and
studied in detail from X-rays to radio wavelengths. With a
measured redshift of z=0.0038 \citep{hvg} (corresponding to a
distance of D=15.2 Mpc for H$_{0}$=75 km~s$^{-1}$ Mpc$^{-1}$), it
provides a scale of only $\sim$ 74 pc/$\arcsec$. A central
nuclear star cluster has an extent of $\sim$ 0.6$\arcsec$
\citep{th97} and a 2.3 kpc stellar bar observed in the near-IR
\citep{sco88,thro89} is surrounded by a circumnuclear starburst
ring.  \citet{te84} found that the infrared emission in NGC~1068
was due to both the Seyfert nucleus (which dominates the 10$\mu$m
emission) and to the star forming regions in the bright $\sim$
3kpc circumnuclear ring (which emits most of the luminosity at
$\lambda>30{\mu}m$). A $Br\gamma$ imaging  study \citep{dav98}
showed a similar morphology and indicated that a short burst of
star formation occurred throughout the circumnuclear ring of
15-16$''$ in radius within the last 4-40 Myr. CO interferometer
observations revealed molecular gas very close to the nucleus
($\sim$0.2$\arcsec$) suggesting the presence of $\sim$
10$^{8}M_{\sun}$ within the central 25pc \citep{schi00}. Recent
high resolution H$_2$ line emission mapping indicates the
presence of two main nuclear emission knots with a velocity
difference of 140 km/s, which, if interpreted as quasi-keplerian,
would imply a central enclosed mass of 10$^{8}M_{\sun}$
\citep{al01}.

In this article, we present the first complete far-infrared
spectrum from 43 to 197$\mu$m showing both atomic and molecular
emission lines (\S 2). We model the composite UV- to far-IR
atomic emission line and continuum spectrum, from our data
and the literature, using photoionization models of both the active
nucleus and the starburst component (\S 3). We also model the mid-
to far-IR continuum emission using a radiative transfer code and
gray body functions for the neutral molecular components (\S 4).
Moreover, two different non-local, non-LTE radiative transfer
codes have been used to model the OH lines (\S 5). Our conclusions
are then given in \S 6.

\section{OBSERVATIONS}

NGC~1068 was observed with the Long Wavelength Spectrometer (LWS)
\citep{cl96} on board the Infrared Space Observatory (ISO)
\citep{kes96}, as part of the Guaranteed Time Programme of the
LWS instrument team. The full low resolution spectrum
(43-197$\mu$m) of NGC~1068 was collected during orbit 605 (July
13, 1997). Two on-source full scans (15,730 seconds of total
integration time) and two off-source (6' N) scans of the
[CII]158$\mu$m line (3,390 seconds of total integration time)
were obtained. On- and off-source scans had the same integration
time per spectral step. Because of the design of the LWS
spectrometer, simultaneously with the 158$\mu$m data, a short
spectral scan of equal sensitivity to the on-source spectrum was
obtained at sparsely spaced wavelengths across the LWS range.

The LWS beam is roughly independent of wavelength and equal to
about 80 arcsec. The spectra were calibrated using Uranus,
resulting in an absolute accuracy better than 30\% \citep{sw96}.
The data analysis has been done with ISAP\footnote{The ISO
Spectral Analysis Package (ISAP) is a joint development by the
LWS and SWS Instrument Teams and  Data Centers. Contributing
institutes are Centre d'Etude Spatiale des Rayonnements (France),
Institute d'Astrophysique Spatiale (France), Infrared Processing
and Analysis Center (United States), Max-Planck-Insitut f\"ur
Extraterrestrische Physisk (Germany), Rutherford Appleton
Laboratories United Kingdom) and the Space Research Organization,
Netherlands.}, starting from the auto-analysis results processed
through the LWS Version 7-8 pipeline (July 1998). To be confident
that newer versions of the pipeline and calibration files did not
yield different results, we have compared our data with the
results obtained using pipeline 10.1 (November 2001) and we did
not find significant differences in the line fluxes or the
continuum.

All the full grating scans taken on the on-source position and the
two sets of data on the off-source position were separately
co-added. No signal was detected in the off-source coadds.  The
emission line fluxes were measured with ISAP, which fits
polynomials to the local continuum and Gaussian profiles to the
lines.  In all cases the observed line widths were consistent
with the instrumental resolution of the grating, which was
typically 1500 km/sec. The integrated line fluxes measured
independently from data taken in the two scan directions agreed
very well, to within 10\%.  The on source LWS spectrum that
resulted from stitching the ten LWS channels together using small
multiplicative corrections in order to match the overlapping
regions of each channel with its neighbors is shown in Fig. 1.
LWS spectra of sources that are very extended within the
instrument beam or that peak off center are typically affected by
channel fringing in the continuum baseline \citep{swi98}.
Fortunately, these spurious ripples are hardly noticeable in our
LWS spectrum, presumably because the far-IR continuum is
centrally concentrated towards the center of the LWS $80''$
beam.

Besides the LWS observations, we also use the SWS observations
presented by \citet{lu00}, to extend the wavelength and
ionization-level coverage. Table~\ref{tbl-1} presents all the ISO
line flux measurements including those from the SWS with their respective
aperture sizes.

\section{THE FINE STRUCTURE LINES}

To be able to better constrain the modeling of the line emission
of NGC~1068, we have combined our far-infrared fine structure line
measurements (Table~\ref{tbl-1}) with ultraviolet, optical and
infrared spectroscopic data from the literature
\citep{kri92,ma96,tho96,lu00}. The complete emission line spectrum
of NGC~1068 from the ultraviolet to the far-IR includes several
low-ionization lines that are primarily produced outside the
narrow line region (NLR) of the active nucleus, as well as
intermediate ionization lines that originate from both starburst
and AGN emission. For this reason, we find that no single model
satisfactorily explains all the observed emission lines. We
identify two main components:

\begin{itemize}
\item[-] an AGN component (the NLR), exciting the high ionization lines
and contributing little to the low-to-intermediate ionization
lines;

\item[-] a starburst component in the circumnuclear ring of
the galaxy \citep[e.g.][]{dav98} that produces the low
ionization  and neutral forbidden lines and some of the
emission in the intermediate ionization lines. This component should
also produce emission associated with photo-dissociation regions (PDRs)
\citep[e.g.][]{kau99}, at the interface with the interstellar
medium of the galaxy.

\end{itemize}

In this section, we will examine separately the two components
that produce the total fine structure emission line spectrum of
NGC~1068, namely the AGN and the starburst, for which we propose
two different computations, and we add together these components
to reproduce the overall observed spectrum from the UV to the
far-IR in \S 3.3.

\subsection{Modeling the AGN}

The first photoionzation model predictions of the mid to
far-infrared emission line spectra of the Narrow Line Regions
(NLR) of active galaxies were presented by \citet{sm92}, well
before the ISO observations could be collected.  \citet{ale00}
used the observed high ionization emission lines to model the
obscured ionizing AGN continuum of NGC~1068 and found that the
best-fit spectral energy distribution (SED) has a deep trough at
4 Rydbergs, which is consistent with an intrinsic ``big blue bump"
that is partially obscured by $\sim$ $6\times10^{19}$ cm$^{-2}$
of neutral hydrogen interior to the NLR. Following their results,
we have simulated their models, although using a different
photoionization code, {\it CLOUDY} \citep[Version
96,][]{fer98,fer00}, and then we have varied the shape of the
ionizing continuum to include the ionizing continuum derived in
\citet{pi94}. Our goal was to test if the \citet{ale00} results
were unique and to fit the remaining emission by a starburst
component, and thereby to derive a composite model of the
complete emission line spectrum of NGC~1068.

Specifically, we explore three plausible AGN SEDs.  Model A
assumes the best fit ionizing spectrum derived by \citet{ale00},
i.e. with a deep trough at 4 Rydberg (log$ f$ = -27.4, -29.0,
-27.4, -28.2 at 2, 4, 8 and 16 Ryd, respectively). An intrinsic
nuclear spectrum of NGC~1068 has also been inferred by
\cite{pi94}. Model B assumes the original ionizing spectrum
derived from \citet{pi94}.  Model C assumes an SED with a {\it
Big Blue Bump} superposed on the \citet{pi94} ionizing continuum
(log$ f$ = -25.8, -25.8, -25.8, -27.4 at 2, 4, 8 and 16 Ryd,
respectively) as expected for the thermal emission of an
accretion disk around a central black hole.  These three AGN
ionizing continua are plotted in Fig.~\ref{fig:ioniz}. For each
of models A, B, and C, we have used two component models with the
same parameters as in \citet{ale00}:  component 1 has a constant
hydrogen density of 10$^{4}$cm$^{-3}$, an ionization parameter
U=0.1, a covering factor c=0.45, a filling factor of 6.5 $\times$
10$^{-3}$ with a radial dependence of the form r$^{-2}$, and
extends from $\sim$ 21 to $\sim$ 109 pc from the center;
component 2 has a density of 2 $\times$ 10$^{3}$ cm$^{-3}$, an
ionization parameter U=0.01, a covering factor of c=0.29, a
filling factor of 6.5 $\times$ 10$^{-4}$ without any radial
dependence, and extends from $\sim$ 153 to $\sim$ 362 pc from the
center. We have also assumed the ``low oxygen" abundances adopted
by \citet{ale00}, in order to be able to compare our results with
theirs\footnote{The adopted gas phase chemical abundances in
logarithmic form are: H: 0.00, He: -1.00, Li: -8.69,  Be: -10.58,
B: -9.21, C:-3.43,  N:-3.96,  O : -3.57,  F : -7.52, Ne: -3.96,
Na: -5.67,  Mg: -4.43,  Al: -5.53, Si: -4.46, P : -6.49, S :
-4.79,  Cl: -6.72,  Ar: -5.60, K : -6.88, Ca: -5.64, Sc: -8.83,
Ti: -6.98,  V : -8.00,  Cr: -6.33, Mn: -6.54, Fe: -4.40, Co:
-7.08, Ni: -5.75, Cu: -7.79, Zn: -7.40}. Because the
grain physics has been updated in the most recent version of CLOUDY
(Version 96), we have included the presence of grains in the models,
using ``Orion-type" grains\footnote{The abundances of the grain
chemical composition, in logarithmic form, are: C: -3.6259,
O: -3.9526, Mg: -4.5547, Si: -4.5547,  Fe: -4.5547}. The inclusion
of grains also allows us to compute the thermal dust continuum
emission from the ionized components (see \S 4).

The inner and outer radii of the emission regions of the two
components, 21, 109, 153 and 362 pc, correspond to angular
distances of about 0.26, 1.4, 1.9 and 4.5 $\arcsec$, respectively.
Table~\ref{tbl-2} reports the predicted line fluxes of the three
AGN models, A, B, and C, together with the observed line fluxes:
the line fluxes are given for each of the two components 1 and 2,
which are treated as independent, and the total flux for each
model is simply the sum of the fluxes of the two components.

We can see from Table~\ref{tbl-2} that only the AGN A and B
models, and not the AGN C model, reproduce most of the observed
high ionization line fluxes.  The low and intermediate ionization
lines, are expected to have partial or full contributions from
starburst and PDR components (see \S 3.2). This first result rules
out the presence of a ``big blue bump" in the ionizing continuum
of NGC~1068. To be able to compare the modeled ultraviolet and
optical lines with the observations, we also listed in
Table~\ref{tbl-2} and ~\ref{tbl-3} their dereddened fluxes,
assuming two values for the extinction: E$_{B-V}$ = 0.4 mag
\citep{ma83} and E$_{B-V}$ = 0.2 mag \citep{ma96}. We find that
the AGN B model overpredicts several of the intermediate
ionization lines, such as [SIV]10.5$\mu$m, [NeIII]15.6$\mu$m and
[SIII]18.7$\mu$m, and this discrepancy increases when adding the
starburst component because these lines are also copiously
produced by that component (see next section). On the other hand,
the [NeII]12.8$\mu$m emission is underpredicted so much so that
even with the inclusion of the starburst component it cannot be
reproduced with this model.  As we discuss further in \S 3.3, a
composite AGN/starburst model using AGN model A reproduces
the [NeII]12.8$\mu$m emission better than the other composite
models.

\subsection{Modeling the starburst ring}

NGC~1068 is known to emit strong starburst emission from the
ring-like structure at a radial distance of $15-16''$ from the
nucleus (total size of $\sim$ 3 kpc), traced for example by the Br
$\gamma$ emission \citep{dav98}. Mid-IR line imaging observations
of NGC~1068 have been published by \citet{lef01} based on ISOCAM
CVF observations. They presented an image of the 7.7 $\mu$m PAH
feature that shows constant surface brightness above the 4th
coutour near the nucleus.  This suggests that star formation is
occuring in the direction of the nucleus so that nuclear spectra
will include some emission from star formation. In the case of
the SWS observations that we are modeling (reported by
\citet{lu00}), three apertures were used at different wavelengths
with the two largest also including portions of the brighter
starburst ring (see Table~\ref{tbl-1}). To estimate how much of
the starburst emission is contained in the different apertures
used in the observations, we have used a continuum subtracted
image in the 6.2 $\mu$m feature produced by C. Dudley (private
communication) using the same ISOCAM CVF data set examined by
\citet{lef01}. The 6.2 $\mu$m feature is more isolated than the
7.7 $\mu$m feature, which is blended with the 8.7 $\mu$m feature
and the silicate absorption feature, but the image compares well
with the published 7.7 $\mu$m image though we have zeroed out the
residuals in a 3$\times$12 arcsec$^{2}$ region centered on the
nucleus.  Based on this image, the SWS 14$\times$20, 14$\times$27
and 20$\times$33 arcsec$^{2}$ slits contain 13, 23 and 46\% of
the 6.2 $\mu$m flux contained in the LWS beam respectively,
without correction for the neglected region of poor residuals
(oriented at 45$^{\circ}$ to our synthetic SWS slits).  Since PAH
features are thought to be a good tracer of PDRs and their
associated startbursts, we adopt these percentages in our model
predictions of SWS line strengths in the starburst models
presented in this section.
In fitting our starburst models to the observations, we
have computed the line fluxes at earth of each centrally
illuminated emitting cloud and then determined the number of clouds
needed to best fit the observed line fluxes.

We have chosen the starburst synthesis modeling program
Starburst99 \citep{leit99}, to produce input ionizing spectral
energy distributions (SEDs) for the {\it CLOUDY} photoionization code.
Models were followed to temperatures down to 50 K to include the PDR
components. We compared the predictions of an instantaneous star formation law
with those of a continuous star formation law. For both types of models we
adopted an age of 5 Myr, a Salpeter IMF ($\alpha$=2.35), a lower
cut-off mass of 1 M$_{\sun}$, an upper cut-off mass of 100
M$_{\sun}$, solar abundances (Z=0.020) and nebular emission
included.  These ionizing SEDs are shown in Fig.~\ref{fig:starb} with
a total mass of M = 10$^{6}$ M$_{\sun}$ for the instantaneous model
and a star formation rate of 1 M$_{\sun}$ yr$^{-1}$ for the continuous
model.  These particular ionizing continuum shapes were selected
because they are consistent with the  Br $\gamma$ equivalent
width observed by \citet{dav98} in the starburst ring. We have
estimated that the Br $\gamma$ equivalent width in each of the
individual regions of the map of \citet{dav98} is in the range
110-180 \AA. According to the \citet{leit99} models (see their
figures 89 and 90), for a value of log(W(Br $\gamma$, \AA))
$\geq$ 2 only instantaneous models with ages less than $\sim$
6$\times$ 10$^{6}$ yrs are allowed.

We report in Table~\ref{tbl-3} the line fluxes predicted for six
different centrally illuminated starburst models, chosing the above instantaneous star
formation SED as the input ionizing continuum and using
{\it CLOUDY} with densities of  n$_H$ = 10, 100, 1000 cm$^{-3}$,
ionization parameters of Log U = -2.5, -3.5 and an
inner cloud radius of 50 pc. As a function of
the adopted density, we then determined the following
numbers of emitting clouds needed to fit the
observations: 33000, 3300, and 330 clouds for the three values of
the density, respectively. The fluxes reported in Table 3 are
therefore the total starburst line fluxes at earth and together
with the nuclear line fluxes of Table 2 can be compared with the
observations.

We have also run models with the continuous star formation law
presented above, but we do not list their results in
Table~\ref{tbl-3}, because the differences in the line flux
predictions compared with the instantaneous models are insignificant,
compared with the effects of density and ionization parameter, as can be seen
from Table~\ref{tbl-3}. This result is not surprising because the
ionizing continua of the two starburst models are quite
similar in shape and the total number of clouds is a free
parameter.
We have also tried continuous starburst models with much
longer ages (10, 20 and 100 $\times$ 10$^{6}$ years) but, because
the shape of the ionizing continuum again does not change
significantly, the resulting emission line spectrum was
indistinguishable from that one derived from the models with an
age of 5 $\times$ 10$^{6}$ years.

In all models the abundances were those typical of HII
regions{\footnote{The adopted gas phase chemical abundances in
logarithmic form are: H: 0.00, He: -1.02, Li: -10.27, Be: -20.00,
B: -10.05, C: -3.52, N: -4.15, O: -3.40, F: -20.00, Ne: -4.22,
Na: -6.52, Mg: -5.52, Al: -6.70, Si: -5.40,  P: -6.80, S: -5.00,
Cl: -7.00, Ar: -5.52, K: -7.96,  Ca: -7.70, Sc:-20.00, Ti: -9.24,
V:-10.00, Cr: -8.00, Mn: -7.64, Fe: -5.52, Co:-20.00, Ni: -7.00,
Cu: -8.82, Zn: -7.70} and grains of
``Orion-type"\footnote{The abundances of the grain chemical composition in
logarithmic form are: C : -3.3249,  O : -3.6516, Mg: -4.2537, Si:
-4.2537,  Fe: -4.2537.} are included. The integration was allowed to
run until the temperature of the gas in the cloud cooled to T=50
K in order to include the photodissociation regions present at
the interfaces of HII regions and molecular clouds.

It is clear from Table~\ref{tbl-3} that the models with the higher
ionization parameter (log U = -2.5) can easily be ruled out,
because their emission in many intermediate ionization lines is
far too high (see e.g. [OIV]26$\mu$m, [OIII]51,88$\mu$m,
[NIII]57$\mu$m). Among the models with the lower
ionization parameter (log U = -3.5), we can exclude model SBR F,
with density n$_H$ = 1000 cm$^{-3}$, because it underestimates
many far-IR lines which are not strongly emitted by the active
nucleus (namely: [SiII]35$\mu$m, [NIII]57$\mu$m, [OIII]88$\mu$m,
[NII]122$\mu$m, [CII]158$\mu$m) while the low density model
(SBR B, with n$_H$ = 10 cm$^{-3}$) overpredicts the [CII]158$\mu$m
line by a factor of 2 relative to the other far-IR lines and
does not reproduce the [OIII] doublet ratio. Finally, the intermediate
density model (SBR D, with n$_H$ = 100 cm$^{-3}$) gives the best
fit to the observed lines, taking into account that the AGN
component must be added to reproduce the total flux as is shown in \S
3.3.

We estimate the average PDR parameters using the models of
\citet{kau99} and the contour plots in \citet{luh03}, the
measured [C II]158 and [O I]145 $\mu$m line fluxes (but not the
[O I]63 $\mu$m line flux which may be affected by absorption
and/or shocks), and the FIR flux integrated over the LWS
spectrum, which we find to be $1.3\times10^{-8}$ ergs cm$^{-2}$
sec$^{-1}$.   Here we assume that the [C II] line emerges
predominantly from PDRs due to the strong starburst, rather than
the diffuse ionized medium.   With this assumption, the average
PDR gas density and UV radiation field are n$_{H_{2}}$ $\sim$
1000 and G$_{0}$ $\sim$ 300 respectively.  We note that if
instead we assume that the [C II] line flux is dominated by the
diffuse ionized medium, using the correction factor estimated by
\citet{mal01}, we obtain a similar gas density n$_{H_{2}}$ $\sim$
1500 but a significantly higher interstellar radiation field
G$_{0}$ $\sim$ 1500.  For both cases, the parameters derived are
in the range of those of the normal galaxies in the \citet{mal01}
sample, consistent with the assumption that most of the FIR flux
originates in the starburst ring.

\subsection{Adding the two components}

Summing the line intensities of each one of the two components,
the composite spectrum of NGC~1068 can be derived and compared
with the observed one. We have chosen three combinations to
compute the composite models, each one with a different AGN
model, while we adopted the starbust model with n$_H$ = 100
cm$^{-3}$ and Log U = -3.5: 1) the first one (that we name CM1,
for Composite Model 1) with the AGN ionizing continuum as
suggested by \citet{ale00} (model AGN A); 2) the second (CM2)
with the original \citet{pi94} (model AGN B); 3) the third (CM3)
with the hypothetical bump (model AGN C). The results of these
three composite models are given in Table~\ref{tbl-4}, compared
to the observed and dereddened values, assuming the two choices
for the extinction (see \S 3.1). We also show the results of the
three composite models in a graphical way in Fig.~\ref{fig:ratio},
where the modeled to the observed flux ratio is given for each
line for the case of an extinction of E$_{B-V}$=0.2 mag.

A simple $\chi^2$ test of the three models resulted in a
reduced $\chi^2$ of 11.6, 17.1 and 177 for the
three models CM1, CM2 and CM3, respectively, for an
extinction of E$_{B-V}$=0.4 mag, while these values become 23, 46
and 325 for E$_{B-V}$=0.2 mag. Thus, of the models explored, CM1
with E$_{B-V}$=0.4 mag provides our best fit to the observations.
We note that model CM1 reproduces the line
fluxes to within a factor of 2.0 on average, with a standard
deviation of 1.3.

\section{THERMAL CONTINUUM SPECTRUM}
\label{sec:thermal}

To model the total mid- and far-infrared thermal dust continuum
of NGC~1068 we have used different computations for each component.  The
modelled emission is shown for the individual components and combined in
Fig.~\ref{fig:cont}. The thermal dust emission from the AGN narrow line
regions and the starburst regions in the ring have been computed using
the \textit{CLOUDY} photoionization models described in the previous sections.
Specifically, the UV continuum reprocessed by the dust present in both
NLR components 1 and 2 of model AGN A has been diluted by the
same covering factors that affect the line emission (c=0.45 and
0.29 for components 1 and 2, respectively).  Similarly we computed the
continuum emission from our best starburst model (SBR D).  While we find
that the continuum produced in this way for the AGN is consistent with
the observed mid-infrared energy distribution (Fig.~\ref{fig:cont}),
accounting for about half of the observed emission at mid-infrared
wavelengths, the emission from dust associated with the starburst ionized
and photodissociated regions,
although similar in shape to the observed continuum, produces only $\sim$ 20\% of
the far-IR continuum. We have indeed performed a search in
parameter space by varying the age of the starburst (from 4 to
6 $\times$ 10$^6$ yrs), the gas density (from 10 to 1000 cm$^{-3}$)
and the radius of the emitting clouds (from 25 to 100 pc), but no
starburst model that could reproduce simultaneously the observed
line and far-infrared continuum emission was found.

If the \textit{CLOUDY} models correctly reflect conditions in both
the ionized and photodissociated gas, these results may imply that the bulk of the starburst far-infrared
continuum arises from dust that is mixed with neutral gas not directly associated with the ionized or photodissociated gas.  However, because photoionization codes such as \textit{CLOUDY} have not been used in the past to model the dust continuum, and because we may not have fully searched parameter space, we are hesitant to over-interpret these results until more detailed comparison with galactic ionized and photoionized regions are carried out.
We have therefore described this starburst thermal dust component,
following \citet{sam02}, in terms of a gray body function with a
temperature of $T=34$ K, and a colder gray body component at
$T=20$ K (see Fig.~\ref{fig:cont}). These are gray body functions with
a steep ($\beta$ = 2) dust emissivity law. Assuming a spherical shell with
radius of 1.5 kpc and thickness of 0.3 kpc, the inferred average H$_2$
molecular density associated with the 34 K component is $3.7$ cm$^{-3}$.
The total mass is $1.2\times10^9$ \Msun\ and $2\times10^9$ \Msun\
for the 34 K and 20 K components, respectively. These estimates
are in reasonable agreement with the $\sim4\times10^9$ \Msun\ derived
for the molecular ring from CO emission \citep{pla91} .

As pointed out above, the dust associated with the ionized NLR
components 1 \& 2 of model AGN A does not quite account for the total
mid-infrared continuum (Fig.~\ref{fig:cont}). Therefore, we have
assumed that the missing mid-infrared arises from the
{\em neutral-nuclear component}, which has been observed in a
variety of molecular lines \citep[e.g.][]{tac94,user04}.
For this neutral component we have assumed a total gas+dust
mass of $M=2\times10^7$ \Msun\ \citep{hel95}, and modelled the expected
continuum using a non-local, spherically symmetric, radiative
transfer code \citep{gon97,gon99}. The molecular nuclear emission
has been resolved into a circumnuclear disk or ring
\citep{schi00}, which here is roughly modelled as a dusty
spherical envelope with inner and outer
radii of 3 and 200 pc, respectively. We assume an AGN luminosity of
$3.7\times10^{10}$  \Lsun, which is a factor of $\sim3$ lower than
the total AGN luminosity, simulating that most of the AGN radiation
escapes through the poles of the molecular disk, and/or is absorbed
in the NLR, and so is not able to heat the molecular gas
\citep[e.g.][]{cam93}. The dust envelope is
divided into a set of spherical shells where the dust temperature
is computed assuming that heating and cooling are equal. We
used a standard silicate/amorphous carbon mixture with optical
constants given by \citet{dra85} and \citet{prei93}. The density
profile was assumed to be $\propto r^{-\beta}$, with $\beta$
regarded as a free parameter.

The resulting mid-infrared emission, obtained with $\beta=1$,
is shown in Fig.~\ref{fig:cont}. The nuclear molecular component
has an averaged H$_2$ density at the inner radius of
$<n_I({\rm H}_2)>=500$ cm$^{-3}$, and a column density
of $N({\rm H}_2)=2\times10^{22}$ cm$^{-2}$.
Once the emission from this neutral component is summed up with
the emission predicted for the NLR components 1 \& 2, a good fit to
the observed emission for $\lambda\ge9$ $\mu$m is obtained.
For $\lambda<9$ $\mu$m, the mid-infrared emission is underestimated,
suggesting the presence of a hot component, probably very close
to the central AGN, which is not included in our models.

\section{THE OH LINES}
\label{sec:ohlines}

\subsection{General remarks}

In NGC~1068, we detect three of the OH rotational lines, all in
emission. As shown in the energy level diagram of
Fig.~\ref{fig:ohlevels}, two of them are fundamental lines,
connecting the ground state $^2\Pi _{3/2}$3/2 level with the
$^2\Pi _{3/2}$5/2 (the in-ladder 119 $\mu$m line) and with the
$^2\Pi_{1/2}$1/2 level (the cross ladder 79 $\mu$m line).  The
third line is the lowest transition of the $^2\Pi_{1/2}$ ladder:
the 163$\mu$m line between the J=3/2 and J=1/2 levels.  The
detected line fluxes are given in Table~\ref{tbl-1}. The fact
that these three lines are {\em all} in emission is in striking
contrast with the OH lines observed in other bright infrared
galaxies, such as Arp~220 \citep{fis99,gon04}, Mrk~231
\citep{har98}, NGC~253 \citep{bra98}, and M~82 \citep{col98}, in
which the 119 $\mu$m fundamental is in absorption. The 79$\mu$m
line is sometimes seen in emission and sometimes in absorption;
the 163$\mu$m line is always seen in emission. In addition to
the detections, the ISO-LWS and SWS observations provide upper
limits on fluxes of the other four lines that arise between the
lowest six rotational levels. The LWS spectra in the vicinity of
the detected lines (and of one of the upper limits), are shown in
detail in Fig.~\ref{fig:ohlines} (histograms). In this section we
discuss the physical conditions necessary to excite these lines,
their probable location within NGC~1068, and detailed model fits
to their fluxes.  A comparison between the observed and modeled
line fluxes is given in Table~\ref{tbl-5} and shown in
Fig.~\ref{fig:ohlines}.

\subsection{The excitation mechanism of the OH lines}

The unique OH emission line spectrum of NGC~1068 can provide a
powerful way to help discriminate between the properties of the
molecular clouds in NGC~1068 and the clouds in other galaxies in
which OH has been observed. Before describing our detailed
radiative transfer calculations, it is instructive to discuss some
conclusions that are model-independent. The emission in the OH
$\Pi_{3/2}\,5/2-3/2$ line at 119 $\mu$m cannot be
explained by absorption of far-infrared photons followed by
cascade down to the upper $\Pi_{3/2}\,5/2$ level of the
transition. Rather, we argue that collisional excitation
dominates. Figure~\ref{fig:ohlevels} shows the energy level
diagram of OH. There are only two possible paths to excite the
119 $\mu$m line via absorption of far-infrared photons: via the
35 $\mu$m and/or the 53 $\mu$m ground-state lines. Excitation by
either of these routes has other observable consequences. In the
case of simple radiative cascading, the Einstein-$A$ coefficients
of the lines involved in the corresponding cascades are such that
if the 35 $\mu$m absorption path were responsible for the
observed 119$ \mu$m line flux, then the OH $\Pi_{1/2}\,5/2-3/2$
line at 98.7 $\mu$m would be approximately 5 times stronger than
the 119 $\mu$m line, while the 98.7 $\mu$m line is not detected.
Hence this possibility is ruled out. Similarly, if absorption in
the 53$\mu$m line were responsible for the observed 119$\mu$m
line flux, then the 163$\mu$m line would be about 5 times stronger
than the 119 $\mu$m line, which it is not. We can therefore
conclude from the constraints provided by the other far-infrared
OH lines that the 119$\mu$m emission line is not the result of
radiative absorption and cascading.  The implication is that OH
excitation through collisions is more important in NGC~1068 than
in the other observed galaxies and therefore that the gas
responsible for the observed emission in the 119$\mu$m line
resides predominantly in relatively dense and warm enviroments in
comparison with these other sources.

The other two observed emission lines,  unlike the 119 $\mu$m
line, need not be collisionally dominated. In the case of the
$\Pi_{1/2}\,3/2-1/2$ 163$\mu$m line, the most likely excitation
mechanism is absorption of photons emitted by dust in the 53 and
35 $\mu$m lines followed by radiative cascade. The upper level of
this transition is 270 K above the ground state
(Fig.~\ref{fig:ohlevels}), so that excitation through collisions
is expected to be ineffective in this line. The excitation
mechanism of the $\Pi_{1/2}-\Pi_{3/2}\,1/2-3/2$ 79 $\mu$m line
could be a mixure of collisional and radiative pumping. The upper
level of this transition is 182 K above the ground state, so that
a warm and dense region could, at least partially, excite the line
through collisions. Nevertheless, the line could be also excited
through the same infrared pumping mechanism that results in the
observed 163 $\mu$m line emission.

In conclusion, the 119 $\mu$m line is collisionally excited, whereas
absorption of photons emitted by dust in the 53 and 35 $\mu$m lines
probably dominates the excitation of the 163 $\mu$m line. The 79 $\mu$m OH line
may in principle be excited through both mechanisms.

\subsection{Constraints on the spatial origin of the 119 $\mu$m OH line}
\label{sec:cons119}

In NGC~1068 two regions with very different physical conditions
can account for the observed OH emission as discussed above: the
compact nuclear region, and the ring and bar where the starburst
is taking place. A warm and dense region is required to account
for the observed 119 $\mu$m line emission, given that the line is
collisionally excited, so the warm and dense neutral nuclear region
around the AGN should be considered a good candidate, despite its
small size \citep[$\sim5''$; e.g.][]{pla91,schi00}, for the
following reasons:

\begin{itemize}
\item[$(i)$] It is warm: there are $\sim10^3$ \Msun\ of hot H$_2$ ($\sim2000$
K) distributed over $\sim5''$ \citep{blie94} that is
thought to be UV or X-ray heated \citep{rot91}. Both PDRs and
XDRs can produce a range of temperatures as high as a few times
$10^3$ K \citep{kau99,stern95,malo96}. From the CO (4-3) to (1-0)
line intensity ratio, \citet{tac94} derive $\sim80$ K for the
bulk of the molecular gas, with a mass of $\sim3\times10^7$
\Msun\ enclosed in within the central $4''$ \citep{hel95}.
\citet{lu00} have reported the detection of pure H$_2$ rotational
lines within the ISO-SWS aperture, and estimated
$\sim2.5\times10^7$ \Msun\ at $\sim200$ K, but these lines may
also arise, at least partially, from the inner regions of the 3
kpc starburst ring.
\item [$(ii)$] The molecular clouds within
the nuclear region are dense, although there is some dispersion in
the values derived by several authors based on HCN emission:
\citet{tac94} derived an H$_2$ density of $\sim10^5$ cm$^{-3}$,
whereas subsequent observations and analysis by \citet{hel95}
yielded a density of $\sim4\times10^6$ cm$^{-3}$. An intermediate
density of $\sim5\times10^5$ cm$^{-3}$ from HCN and CS, and lower
for other tracers, has been recently derived by \citet{user04}.
\item [$(iii)$] The
OH abundance is expected to reach high values in regions
exposed to strong incident UV fields \citep[PDRs,][]{stern95},
and in particular in X-ray dominated regions
\citep[XDRs,][]{lepp96}. The remarkable chemistry found by
\citet{user04} in the circumnuclear disk of NGC~1068 is
indicative of an overall XDR and suggests a high OH abundance in
the nuclear region.
\end{itemize}

Given that the OH 119$\mu$m line is collisionally excited, the
possibility that the line might arise from the nuclear region can be checked
by computing the amount of warm gas required to account for the observed
emission:
\begin{equation}
M_{w}({\rm M_{\odot}})=1.6\times10^7 \times \left[
\frac{10^{-5}}{X({\rm OH})} \right] \times \left[
\frac{5\times10^{5}\,{\rm cm^{-3}}}{n({\rm H}_2)} \right] \times
\left[ \frac{4.3\times10^{-11}\,{\rm cm^3\,s^{-1}}}{<c_{lu}>}
\right], \label{eq:masswarm}
\end{equation}
where $X({\rm OH})$ is the OH abundance relative to H$_2$, and
$<c_{lu}>$ is the collisional rate for excitation from
the ground $\Pi_{3/2}\,3/2$ level to the $\Pi_{3/2}\,5/2$ one.
Equation~\ref{eq:masswarm} assumes that, although the line could
be optically thick, it is effectively optically thin, and makes
use of the observed flux of $1.2\times10^{-12}$ erg s$^{-1}$
cm$^{-2}$. The reference value for the collisional rate,
$<c_{lu}>=4.3\times10^{-11}$ cm$^3$ s$^{-1}$, corresponds to gas
at 80 K \citep*{offer94}; it decreases by a factor of
$\approx2.7$ for gas at 50 K and increases by a factor of 3 for
gas at 200 K.

The reference OH abundance we use in this estimate, $10^{-5}$, is
the result of two separate studies: first, calculations of
molecular abundances by \citet{lepp96} have shown that the OH
abundance in XDRs is expected to be about two orders of magnitude
higher than the abundance of HCN and HCO$^+$. The authors in fact
suggested the possibility that the high HCN/CO ratio observed in
the nuclear region of NGC~1068 could be a consequence of enhanced
X-ray ionization. Second, the possibility of a chemistry
dominated by X-rays has found support from observations by
\citet{user04}, who derive abundance ratios of HCN, HCO$^+$, and
CN in general agreement with predictions for XDRs. Since the HCN
abundance derived by \citet{user04} is $\sim10^{-7}$, $X$(OH) in
XDRs could attain values as high as $10^{-5}$. On the other hand,
the density of $5\times10^5$ cm$^{-3}$ derived by \citet{user04}
has been adopted as the reference value in eq.~\ref{eq:masswarm}.
Finally,
the mass of molecular gas derived from the emission of several
molecular tracers is expected to be at least $\sim2\times10^7$ \Msun\
\citep{hel95}, which is similar to the value required in
eq.~\ref{eq:masswarm}. From
these estimates we conclude that, if the OH abundance is as high
as $\sim10^{-5}$ (i.e. if the predictions for XDRs are applicable
to the nuclear region of NGC~1068), the bulk of the OH 119 $\mu$m
line could arise there. This possibility would naturally explain
why NGC~1068 is unique in its 119 $\mu$m line emission among
galaxies with full LWS spectra.

Finally we ask whether the OH 119$\mu$m line could arise from an even
more compact region, i.e., from a torus with a spatial scale of 1 pc
surrounding the central AGN. According to typical parameters given by
\citet{kro89}, a torus is expected to be
hot ($\sim10^3$ K), could have densities of $10^7$ cm$^{-3}$, and
therefore a mass of $\sim10^5$ \Msun. Also, the OH abundance is expected
to be very high, $5\times10^{-5}-10^{-4}$. Eq.~\ref{eq:masswarm} shows
that the relatively low mass of the torus (about 2 orders of magnitude
lower than the entire nucleus) could be compensated by the
the higher density, OH abundance, and temperature expected there, so that
this possibility cannot be neglected.

The reference values for the nuclear abundance and density
adopted in eq.~\ref{eq:masswarm} are rather uncertain (and
possibly extreme). The continuum models of \S 4 indicate that the
mass associated with the 34 K dust component, which is
identified with the starburst ring, is $1.2\times10^9$
\Msun. If $\sim5$\% of this mass corresponds to warm molecular
gas rich in OH, the amount of extended warm gas is
$\sim6\times10^7$ \Msun. According to eq.~\ref{eq:masswarm}, the
OH emission at 119 $\mu$m can then also be explained as arising
in the ring if the associated PDRs, with assumed OH abundance of
$2\times10^{-6}$ \citep{stern95,goi02,gon04}, have densities of
${\rm a\,\, few}\times10^5$ cm$^{-3}$.  Since \citet{psb99} found
that most of the extended molecular gas resides in dense, compact
clouds, this scenario seems also possible. However, the continuum
from the starburst at 119 $\mu$m is strong, so that one expects
that eq.~\ref{eq:masswarm} is in this case underestimating
$M_{w}$, and the quoted physical parameters, $X$(OH) and $n({\rm
H}_2)$, are lower limits. The effect of dust emission is
discussed in detail below.

In conclusion, a definitive answer to the issue of the spatial origin
of the OH 119 $\mu$m emission cannot be inferred from only the flux observed
in the 119 $\mu$m line. Nevertheless, useful constraints on this subject are
given: the line could be either explained as arising from the nucleus, with
a required OH abundance $\sim10^{-5}$, or from
the extended ring, with OH abundance $>2\times10^{-6}$ and density
$>{\rm a\,\, few}\times10^5$ cm$^{-3}$. Nevertheless, the radiative transfer
models described below, which take into account the effect of the continuum
emission and the excitation of the 79 and 163 $\mu$m lines, point towards a
nuclear origin of the OH emission.

\subsection{Outline of the models}

Analysis of the OH 79 and 163 $\mu$m lines requires the use of
detailed radiative transfer calculations since, as pointed out
above, the emission in these lines is expected to be strongly
influenced by absorption of far-infrared continuum photons. We
therefore proceeded to model the OH lines with two different
codes, and confirmed that the results were in good agreement with
each other. One of them, described in \citet{gon97,gon99}, has
been recently used to model the far-infrared spectrum of Arp 220
\citep{gon04}. The other is a Monte Carlo radiative
transfer code used as part of a detailed study of all the OH
lines observed by ISO in galaxies \citep{smi04a,smi04b}. The code
was developed for the Submillimeter Wave Astronomy Satellite
(SWAS) mission and is a modified and extended version of the
Bernes code \citep{ber79} but which includes dust as well as gas
in the radiative transfer, and also corrects some optical depth
calculations from the original code \citep{ash00}. Both methods
are non-local, non-LTE, assume spherical symmetry, and include a
treatment of continuum photons from dust mixed in with the gas.
Also, both codes take input as a series of concentric shells,
each of which is assigned a size, gas and dust temperature, H$_2$
density, velocity and turbulent velocity width, and molecular
abundance relative to H$_2$. The statistical equilibrium
populations of OH in each spherical shell are computed by
including the excitation by dust emission, excitation through
collisions, and effects of line trapping. We ran two models to
simulate the nucleus of the galaxy and the starburst extended
ring, described in \S 5.4.1 and 5.4.2.

\subsubsection{Models for the nuclear emission: constraints on the spatial
origin of the 79 and 163 $\mu$m OH lines}

We present models for the nuclear OH emission
that {\em implicitly assume that the 119 $\mu$m emission line arises
from the nuclear region}: $X$(OH)$=10^{-5}$ is adopted, as well
as densities $\sim5\times10^5$ cm$^{-3}$ for the bulk of the
emitting gas. By assuming a pure nuclear origin for
the 119 $\mu$m line, we check whether the other two OH lines
could, in such a case, arise from the same nuclear region or require
a more extended spatial origin.

The models use the dust parameters derived for the
nuclear molecular region described in \S 4. The possible
contribution to the OH excitation of far-infrared photons arising
from the components 1 \& 2 of model AGN A is ignored, i.e. only the dust
coexistent with the molecular gas is taken into account, with
predicted flux of $24-27$ Jy at 35--53 $\mu$m
(Fig.~\ref{fig:cont}). The gas temperature is assumed uniform and
equal to 70 K \citep[e.g.][]{tac94}.


The H$_2$ densities derived from the dust model (i.e. a peak
density of $<n_I({\rm H}_2)>$ = 500 cm$^{-3}$ at the
inner radius) are not compatible with the densities inferred from
different molecular tracers. This indicates that the medium is
extremely clumped, as has been also argued elsewhere
\citep[e.g.][]{cam93,tac94}. In order to account approximately
for this clumpiness in our models, the following strategy is
adopted: we use the ``real'' $n({\rm H}_2)\sim5\times10^5$
cm$^{-3}$ values for the bulk of the gas, and compute the volume
filling factor $f_v=<n({\rm H}_2)>/n({\rm H}_2)$, where the
average value is that inferred from the dust model. The expected
abundances of OH and the dust relative to H$_2$, $X$(OH) and
$X$(dust), are then multiplied by $f_v$, so that the right OH and
dust column densities are used in the calculations together with
the right density values. The same density profile $r^{-1}$ that
was used in the dust model is adopted, so that $f_v$ is uniform
throughout the nuclear region.

The modeled fluxes are convolved with the ISO-LWS grating
resolution and are compared with the data in
Fig.~\ref{fig:ohlines}. Solid black lines show the results for the
nuclear model that assumes $X$(OH)=$10^{-5}$ and
$f_v=2\times10^{-5}$, the latter value implying a density in the
outer regions (where the bulk of the emission is generated) of
$5\times10^5$ cm$^{-3}$. The radial OH column density is
$N({\rm OH})=2\times10^{17}$ cm$^{-2}$. Besides the 119 $\mu$m
line, the model nearly reproduces the emission in the 79
and 163 $\mu$m lines and is consistent with the upper limits
given in Table~\ref{tbl-5}.

We also checked the excitation mechanism of the other OH lines by
generating an additional model with the same parameters as above
except for the continuum emission, which is now turned off. In
this model, therefore, the lines are excited exclusively through
collisions with H$_2$. The resulting flux of the 119 $\mu$m line
remains unchanged when the dust emission is ignored, confirming
that the line is collisionally excited. On the other hand, the
flux densities of the 79 and 163 $\mu$m lines decrease in the
``pure-collisional'' model by factors of 2 and 6, respectively,
showing that the emission in these lines is much more affected by
radiative pumping. We conclude that, if the OH abundance in the
nucleus were high enough to account for the collisionally excited
119 $\mu$m line, the observed fluxes in the 79 and 163 $\mu$m
lines can also be explained as arising in the same nuclear region.

\subsubsection{Models for the starburst emission}

Two simple different approaches have been used to model the OH
emission from the starburst. First, we have roughly modeled the
whole starburst region as a spherical shell with external radius
of 1.5 kpc, thickness of 0.3 kpc, and average H$_2$ density
$<n_I({\rm H}_2)>$ = 3.7 cm$^{-3}$, so that the corresponding
continuum emission is reproduced with $T_d=34$ K
(section~\ref{sec:thermal}). As shown in
section~\ref{sec:cons119}, the OH 119 $\mu$m emission requires
densities of ${\rm a\,\, few}\times10^5$ cm$^{-3}$, so that we
have assumed a volume filling factor $f_v=7.5\times10^{-6}$
and therefore a ``real''
density $n({\rm H}_2)=5\times10^5$ cm$^{-3}$. The
kinetic temperature is assumed to be $T_k=100$ K, and the OH
abundance is $X$(OH)$=2\times10^{-6}\times f_h$, where $f_h=0.05$
is the assumed fraction of warm gas. The result of this model is
shown in Fig.~\ref{fig:ohlines} (upper dotted lines). The 119
$\mu$m line is reproduced, but the flux densities of both the 79
$\mu$m and 163 $\mu$m lines are strongly underestimated. The
reason is that the model implicitly assumes that the continuum
emission, responsible for the excitation of those lines, arises
from a very large volume, so that the {\em radiation density} is
weak and has negligible effect on the line excitation.

Since the OH emission is expected to arise from compact, discrete
PDRs in the vicinity of O or early B stars, where the continuum
infrared radiation density is expected to be stronger than assumed
above, we have also tried an
alternative approach, which consists of modelling an individual
``typical'' cloud of the starburst. We first model the continuum from
an individual cloud by assuming a central heating source and computing
the dust equilibrium temperatures at each radial position that result from
the balance of heating and cooling. The continuum model is adopted if
$(i)$ leaving aside a scaling factor ($N_c$, the number of clouds in the
ensemble), the resulting SED is similar to that of the far-infrared emission
of the starburst (i.e. the 34 K component found in
section~\ref{sec:thermal}); the value of $N_c$ is determined by
requiring that the absolute continuum flux from the ensemble of clouds
is equal to that observed for the 34 K component;
and $(ii)$ we require that the total mass of the ensemble does not exceed
the mass inferred from the non-nuclear region ($<4\times10^9$ \Msun).
Once the continuum is fitted, calculations for OH are performed by assuming
$T_k=T_d$, and $X$(OH)$=2\times10^{-6}$.

Several models with various density profiles were found to match
the above two requirements. The common characteristic of all of
them is the relatively high column density of the individual
clouds, $N({\rm H_2})>10^{23}$ cm$^{-2}$, which is a consequence
of the low effective dust temperature (34 K) of the SED. The
results of the simplest model, characterized by a flat density
profile, $n({\rm H_2})=5\times10^5$ cm$^{-3}$, are given here for
reference. With a radius of $5\times10^{17}$ cm, a stellar
luminosity of $2\times10^4$ \Lsun, and $N_c=4\times10^6$, the
resulting SED is similar to that of the 34 K component (\S
4). For these clouds we obtain a total mass of $1.8\times10^9$
\Msun. The predicted OH emission/absorption is shown in
Fig.~\ref{fig:ohlines} (lower dashed lines). In spite of the
relatively high density and temperatures (28-250 K) throughout
the cloud, the 119 $\mu$m line is predicted to be too weak, and
the 79 $\mu$m line is predicted in absorption. We have found this
result quite general: in models where the OH abundance is high
enough and the radiation density becomes strong enough to pump
the 163 $\mu$m emission, the continuum at 79 and 119 $\mu$m is
absorbed by OH and the predicted emission in the corresponding
lines is reduced. Models that assume a density profile of
$r^{-1}$ generally predict the 119 $\mu$m line in absorption. In
some models where the OH abundance was allowed to vary with
radial position, the 79 $\mu$m line was predicted in emission but
by far too weak to account for the observed flux density.

In conclusion, no starburst model is found to reproduce
satisfactorily the emission observed in the three OH lines. If
the local infrared radiation density is strong enough to pump the
163 $\mu$m line, the other two OH lines are expected to be weak
or in absorption. Furthermore, the high density assumed for the
starburst region would produce a relatively high HCN/CO intensity
ratio, which is on the contrary $\sim0.01$ in the spiral arms
\citep{hel95}. Finally, the PDR models described in \S 3.2
indicate a density of $1-1.5\times10^3$ cm$^{-3}$, i.e. a density
much lower than that required to account for the flux density of
the OH 119 $\mu$m line. Therefore, and despite the simplicity of
our models, taken together the analysis of the OH lines and the
derived PDR parameters indicate that the bulk of the OH emission
arises from X-ray dominated nuclear regions.

\section{CONCLUSIONS}

The main results of this article can be summarized as follows:

\begin{itemize}

\item The complete far-infrared (50-200$\mu$m) spectrum of NGC~1068 has been
observed for the first time. The far-infrared ISO-LWS spectrum
has been complemented with the mid-infrared data of ISO-SWS and
with shorter wavelength (UV, optical and near-IR) data from the
literature to assemble a composite atomic spectrum as complete as
possible with the aim of modeling the different line emission
components at work. This approach has been necessary especially
because of the poor spatial resolution of the ISO spectrometers,
which were not able to spatially separate the emission
components. The lines have been interpreted as arising from two
physically distinct components: the AGN component and a starburst
component, the first one nuclear and the second one located in
the ring at a radius of 15-16 $\arcsec$ from the nucleus.
Both components are characterized by the presence of dust grains,
producing strong continuum emission in the mid- and far-infrared.
The density and ionization parameter of the $\sim$
5 $\times$ 10$^6$ year old starburst are found to be n$_H$ $\sim$
100 cm$^{-3}$ and log U = -3.5, respectively. Three composite
models have been computed with different AGN components: the
first one has the ionizing continuum as derived from
\citet{ale00}, showing a deep trough at energies of a few
Rydberg; the second has the monotonically decreasing ionizing
continuum given by \citet{pi94} and the third has a ``big blue
bump". Two values of the visual extinction ($E_{B-V}=0.2$ and
0.4) have been adopted to correct the optical and ultraviolet
line fluxes for the reddening. The agreement between the
composite model with an AGN ionizing continuum characterized by
the deep trough suggested by \citet{ale00} is very satisfactory,
taking into account both the simplicity of the photoionization
models chosen to avoid dealing with too many free parameters and
the large number of lines which originate in different physical
regimes. The agreement between the observed spectrum and what is
predicted using the canonical ionizing continuum is
slightly poorer, while the presence of a big blue bump
is ruled out.

\item The 50-200$\mu$m continuum has been modeled using
different components arising from both the nucleus and the
starburst ring. For the nucleus, we have combined the dust
emission from the ionized components in the narrow line regions
modeled by \textit{CLOUDY} with the neutral component
reproduced by the radiative transfer code used for the OH
molecular emission. For the starburst ring, our \textit{CLOUDY} modelling of the
ionized + PDR components could not reproduce
the far-infrared emission, while similtaneously fitting the far-IR lines.
Instead we fit the observed continuum by a neutral molecular
component, reproduced by two gray body components at temperatures
of 20K and 34K, assuming a steep ($\beta$=2) dust emissivity law.

\item The unique OH emission in the 119 $\mu$m line cannot be explained
in terms of OH excitation through absorption of 35 and 53 $\mu$m
photons emitted by dust, but rather it is collisionally excited.
This indicates the presence of a warm and dense region with high
OH abundance. A simple excitation analysis yields two main
alternatives for the {\em spatial origin} of the observed 119
$\mu$m line emission: $(i)$ the nuclear region, with $2\times10^7$
M$_\odot$ of warm gas (80 K), an average density of $n({\rm
H}_2)=5\times10^{5}$ cm$^{-3}$, and an OH abundance of
$\sim10^{-5}$; $(ii)$ the starburst region, if $\sim5$\% of the
associated mass ($\sim6\times10^7$ M$_\odot$) is warm ($\sim100$
K), dense (${\rm a\,\, few}\times10^{5}$ cm$^{-3}$), and rich in
OH ($X$(OH)$\sim2\times10^{-6}$).

\item Radiative transfer models that simulate the emission/absorption
in all the OH lines have been performed for both the nuclear and
the starburst region. The models for the nucleus quantitatively
account for the emission in the three OH lines if the nuclear
physical conditions pointed out above are assumed. On the other
hand, no starburst model is found to match the three OH lines
simultaneously, because the strong far-infrared continuum tends to
produce absorption, or to weaken the emission, in the OH 119 and
79 $\mu$m lines (as observed in other galaxies). Therefore,
although some contribution from the extended starburst cannot be
ruled out, our models indicate that the bulk of the OH emission
arises in the nuclear region. The high nuclear OH abundance
required to explain the emission strongly suggest a chemistry
deeply influenced by X-rays, i.e., an {\em X-ray dominated
region}.
\end{itemize}

\acknowledgments The authors acknowledge the LWS Consortium, lead
by Prof. Peter Clegg, for having built and operated the LWS
instrument and solved many instrumental and data reduction
problems. We acknowledge discussions with Dr. Chris Dudley
and thank him for reduction and analysis of the ISOCAM 6.2 $\mu$m
image that we used in this work.  We also thank Dr. Matt Ashby for
his help with the Monte-Carlo modeling of the OH lines.
The ESA staff at VILSPA
(Villafranca, Spain) is also acknowledged for the ISO mission
operational support. HAS acknowledges support from NASA Grant
NAG5-10659; E.G-A would like to thank the Harvard-Smithsonian
Center for Astrophysics for its hospitality while he was in
residence during this research. JF acknowledges support from the
NASA LTSA program through contract S-92521-F and from the Office of
Naval Research.

\clearpage

\begin{figure}
\plotone{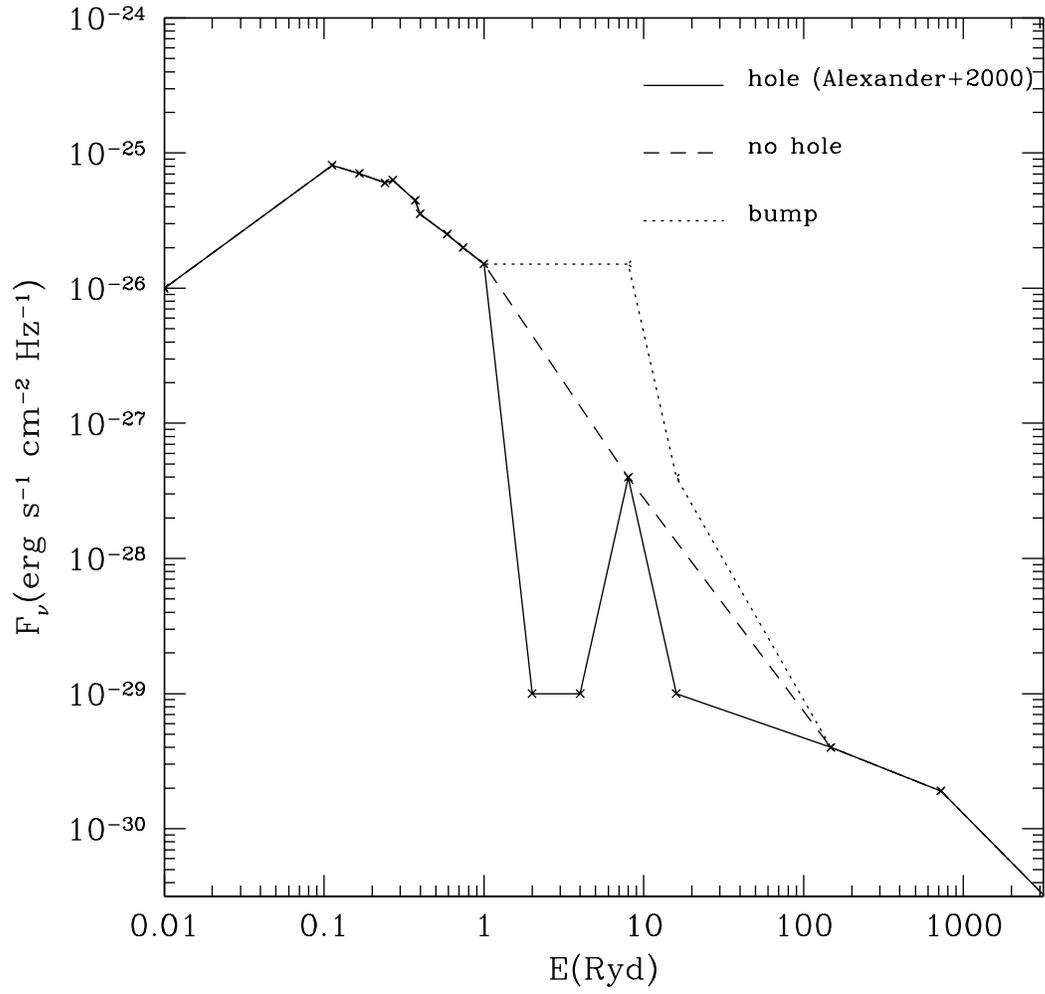} \caption{The AGN ionizing continua used as input
for the photoionization models of NGC~1068. The three continua
differ in the frequency region between 1$>$E$_{Ryd}$$>$100, while
outside this region the \citet{pi94} spectrum was adopted. The
solid line shows the continuum derived from \citet{ale00}; the
dashed line shows a simple power law interpolation; the dotted
line shows the presence of the predicted "big blue bump".
\label{fig:ioniz}}
\end{figure}

\begin{figure}
\plotone{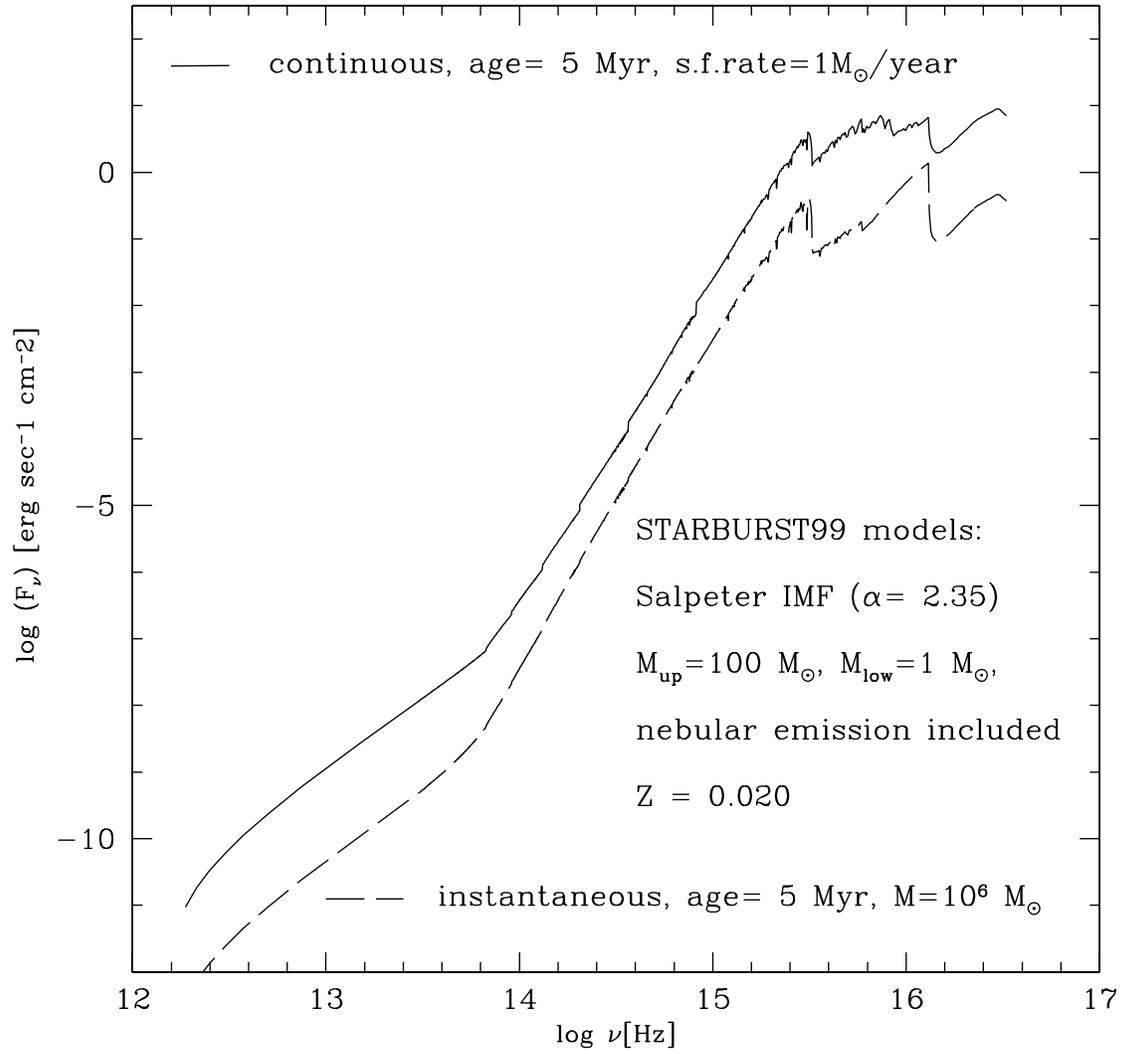} \caption{The starburst spectral energy distributions
used as input for the photoionization models of NGC~1068. The two
continua are taken from \cite{leit99} and represent a continuous
starburst model (solid line) and an instantaneous model (broken
line), both with ages of 5 Myr. \label{fig:starb}}
\end{figure}

\begin{figure}
\plotone{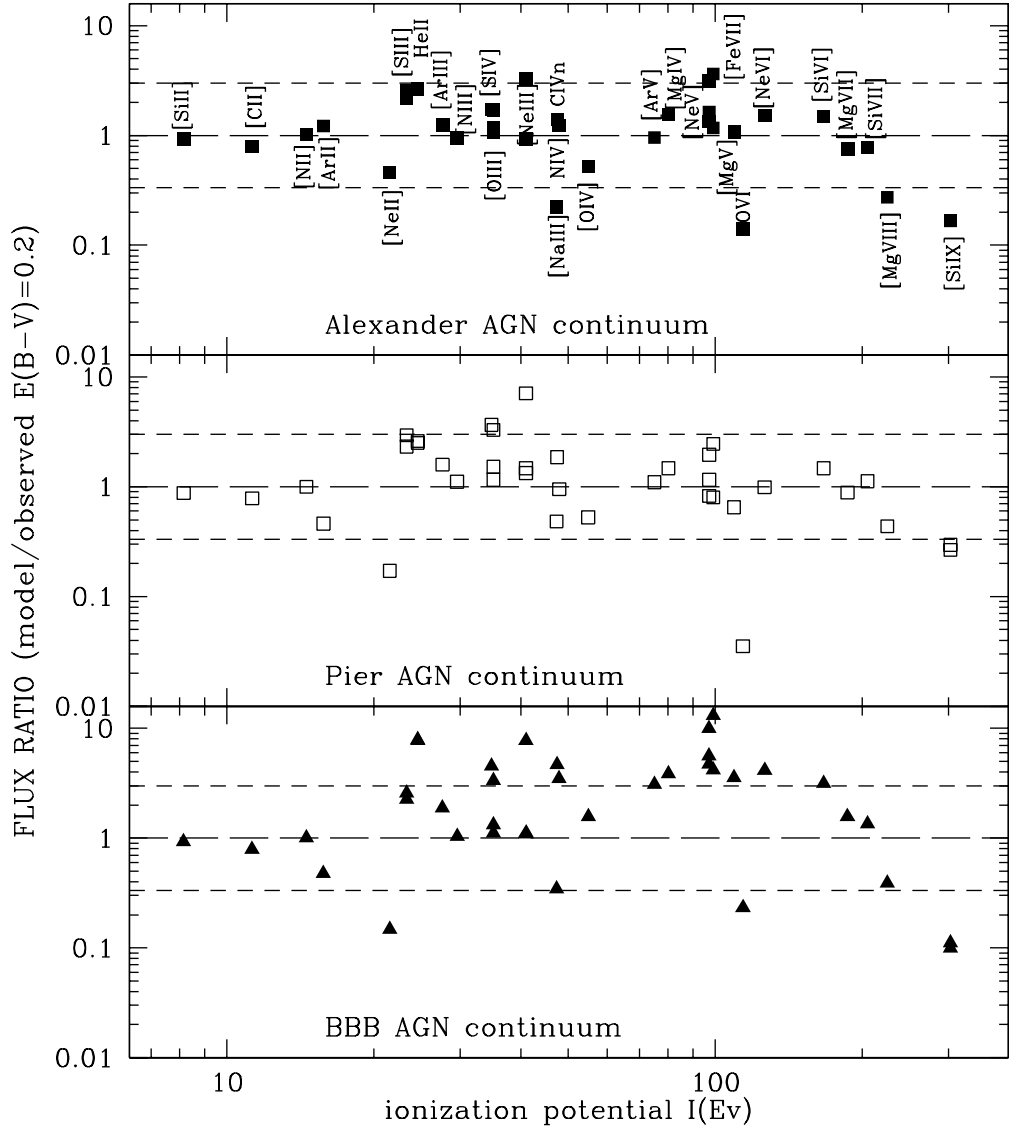} \caption{The comparison of the composite models
with the observations is shown as the ratio of modeled to
observed flux ratio for each line, with the ionization potential
in the x-axis. The assumed reddening is E(B-V)=0.2. Panels from
top to bottom: model CM1, CM2 and CM3. The short dashed lines
represent flux ratios within a factor 3 either ways.
\label{fig:ratio}}
\end{figure}

\begin{figure}
\epsscale{0.75} \plotone{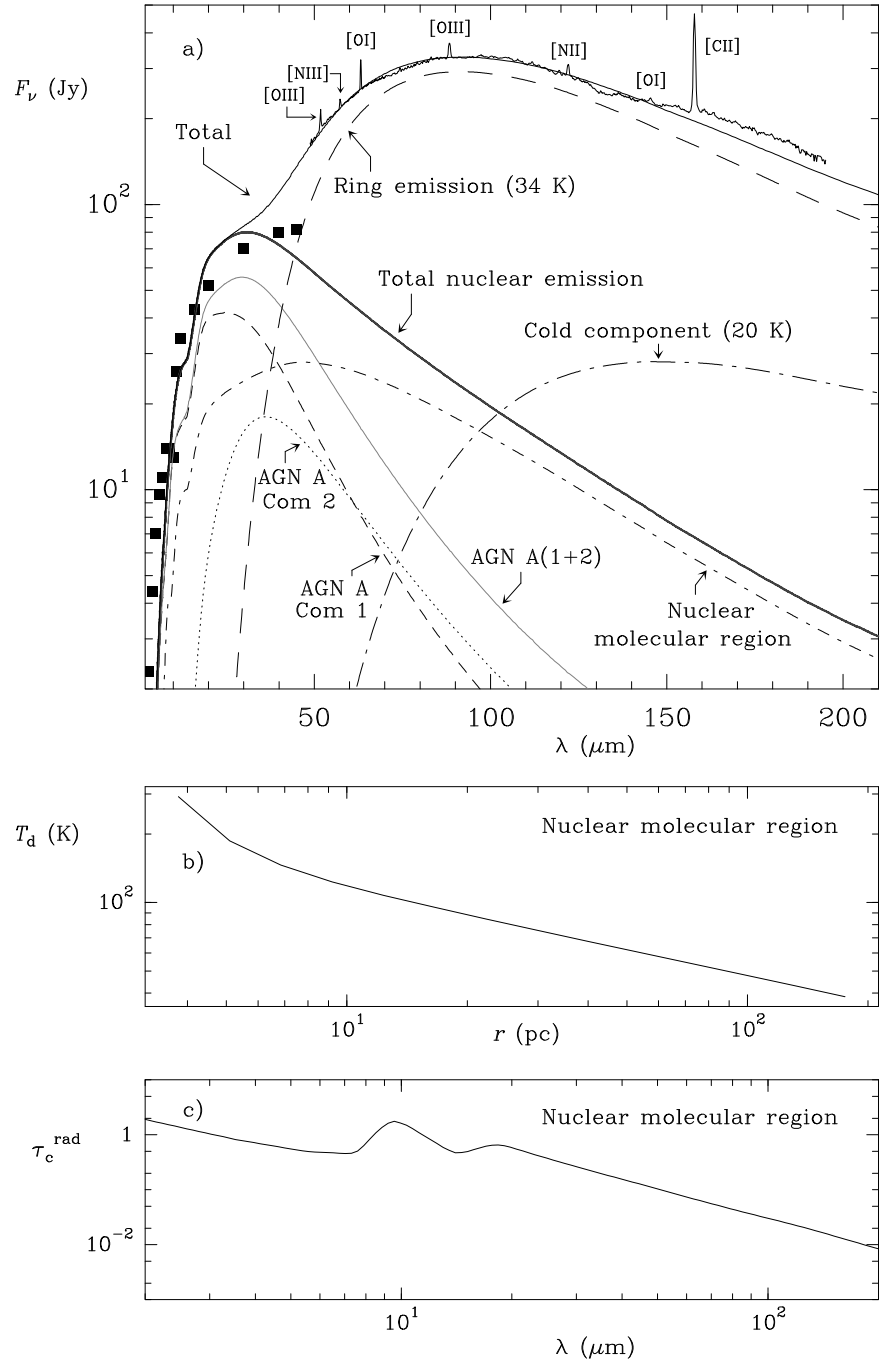} \caption{a) Spectral energy
distribution of NGC~1068 and model fit. ISO-SWS fluxes are taken
from \citet{lu00}. The model fit is the composition of $(i)$ the
NLR components 1 \& 2 of model AGN A, $(ii)$ the emission from the
molecular nuclear region, $(iii)$ the 34 K starburst emission, and
$(iv)$ the cold 20 K component. b) and c) Dust temperature versus
the radial position and radial continuum opacity versus wavelength
for the nuclear molecular region.
\label{fig:cont}}
\end{figure}

\begin{figure}
\epsscale{0.55} \plotone{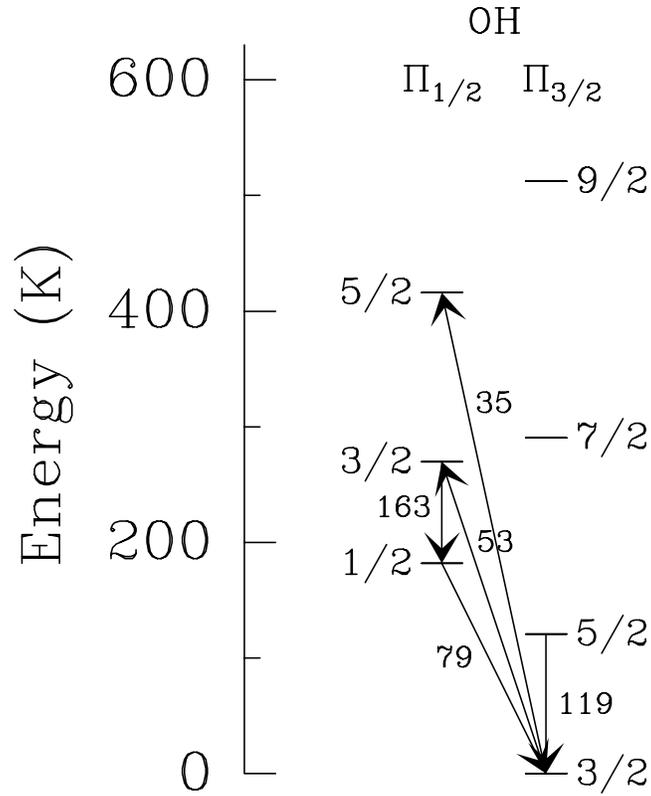} \caption{Energy level diagram of
OH. Rotational levels with energies up to 600 K are shown; the
three lines detected in NGC 1068 are indicated with arrows, as
well as the 35 and 53 $\mu$m lines that could play an important
role in the radiative excitation. The wavelengths are indicated
in $\mu$m. $\Lambda$-doubling is ignored because the
$\Lambda$-doublets are not resolved with the ISO grating
resolution. \label{fig:ohlevels}}
\end{figure}

\begin{figure}
\epsscale{0.55} \plotone{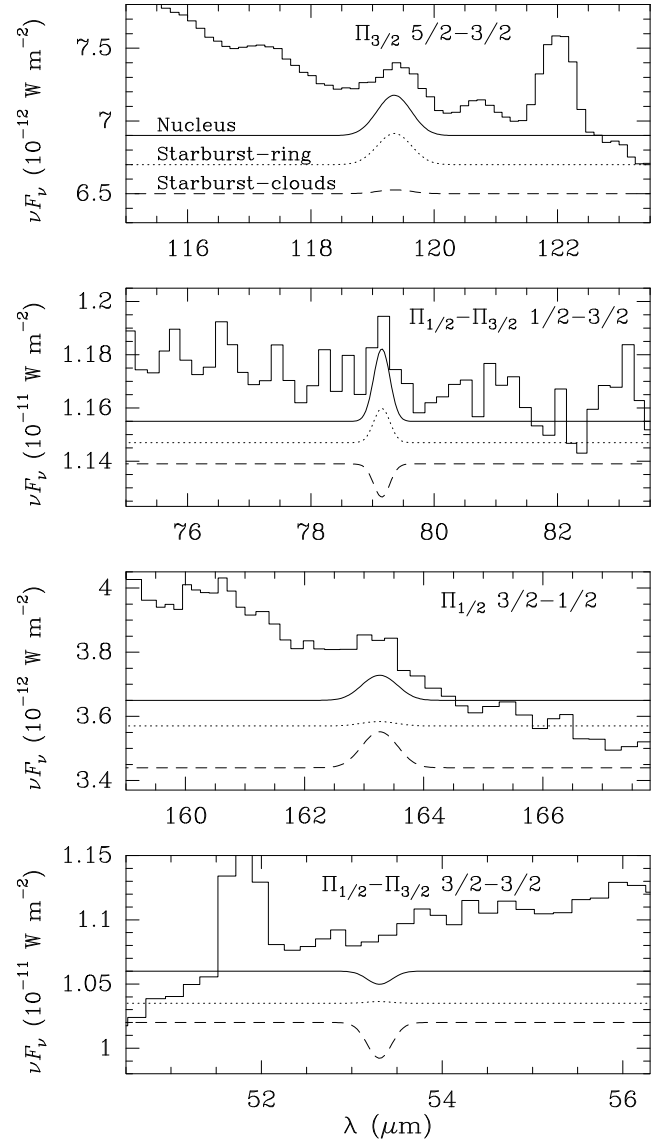} \caption{Comparison between the
observed OH lines and model results. As indicated in the upper
panel, the upper modeled spectrum (solid lines) corresponds to
the model for the nucleus, the middle one (dotted lines)
corresponds to the starburst modelled as a whole, and the lower
one (dashed lines) corresponds to the starburst modelled as an
ensemble of individual clouds (see text for details).
\label{fig:ohlines}}
\end{figure}

\begin{deluxetable}{lcccc}
\tabletypesize{\scriptsize}
\tablecaption{Measured line fluxes
from the LWS and SWS grating spectra, with 1$\sigma$
uncertainties. \label{tbl-1}}

\tablewidth{0pt} \tablehead{ \colhead{Line}
& \colhead{$\lambda$}
& \colhead{Flux} & \colhead{Aperture} &
\colhead{reference}\\
\colhead{} & \colhead{($\mu$m) } &
\colhead{$(10^{-13}~erg~s^{-1}~cm^{-2})$} &
\colhead{$(\arcsec^{2})$} & \colhead{} }
\startdata
${\rm [Si~IX ]}$ $^{3}P_{2}\rightarrow {^{3}P_{1}}$      & 2.584 & 3.0  &
14 $\times$ 20 & 1 \\
${\rm [Mg~VIII]}$ $^{2}P_{3/2}\rightarrow {^{2}P_{1/2}}$ & 3.028 &  11.
$\pm$ 1.1  & 14 $\times$ 20 & 1 \\
${\rm [Si~IX ]}$  $^{3}P_{1}\rightarrow {^{3}P_{0}}$     & 3.936 &  5.0
$\pm$ 0.6  & 14 $\times$ 20 & 1 \\
${\rm [Mg~IV]}$ $^{2}P_{1/2}\rightarrow {^{2}P_{3/2}}$   & 4.487 & 7.6
$\pm$ 1.5  & 14 $\times$ 20 & 1\\
${\rm [Ar~VI]}$ $^{2}P_{3/2}\rightarrow {^{2}P_{1/2}}$   & 4.529 & 15.
$\pm$ 3.  & 14 $\times$ 20 & 1\\
${\rm [Fe~II]}$ $^{a4}F_{9/2}\rightarrow {^{a6}D_{9/2}}$ & 5.340 &
5.0     & 14 $\times$ 20 & 1\\
${\rm [Mg~VII]}$ $^{3}P_{2}\rightarrow {^{3}P_{1}}$      & 5.503 & 13.  &
14 $\times$ 20 & 1\\
${\rm [Mg~V]}$ $^{3}P_{1}\rightarrow {^{3}P_{2}}$        & 5.610  & 18.
$\pm$ 2.  & 14 $\times$ 20 & 1\\
${\rm [Ar~II]}$ $^{2}P_{1/2}\rightarrow {^{2}P_{3/2}}$   & 6.985  & 13.   &
14 $\times$ 20 &  1\\
${\rm [Na~III]}$ $^{2}P_{1/2}\rightarrow {^{2}P_{3/2}}$  & 7.318 & 5.8   &
14 $\times$ 20 &  1\\
${\rm [Ne~VI]}$ $^{2}P_{3/2}\rightarrow {^{2}P_{1/2}}$   & 7.652  & 110.
$\pm$ 11.  & 14 $\times$ 20 &  1\\
${\rm [Fe~VII]}$ $^{3}F_{4}\rightarrow {^{3}F_{3}}$      & 7.815  & 3.0   &
14 $\times$ 20 &  1\\
${\rm [Ar~V]}$ $^{3}P_{2}\rightarrow {^{3}P_{1}}$        & 7.902 & $<$ 12.
& 14 $\times$ 20 &  1\\
${\rm [Na~VI]}$ $^{3}P_{2}\rightarrow {^{3}P_{1}}$       & 8.611 & $<$ 16.
& 14 $\times$ 20 &  1\\
${\rm [Ar~III]}$ $^{3}P_{1}\rightarrow {^{3}P_{2}}$      & 8.991  & 23.0
$\pm$ 3.3  & 14 $\times$ 20 & 1\\
${\rm [Fe~VII]}$ $^{3}F_{3}\rightarrow {^{3}F_{2}}$      & 9.527  &
4.0    & 14 $\times$ 20 & 1\\
${\rm [S~IV]}$ $^{2}P_{3/2}\rightarrow {^{2}P_{1/2}}$    & 10.510 & 58.
$\pm$ 6.  & 14 $\times$ 20 &  1\\
${\rm [Ne~II]}$ $^{2}P_{3/2}\rightarrow {^{2}P_{1/2}}$   & 12.813 &
70.    & 14 $\times$ 27 & 1 \\
${\rm [Ar~V]}$ $^{3}P_{1}\rightarrow {^{3}P_{0}}$        & 13.102 & $<$ 16.
& 14 $\times$ 27 &  1\\
${\rm [Ne~V]}$ $^{3}P_{2}\rightarrow {^{3}P_{1}}$        & 14.322 & 97.
$\pm$ 9.7  & 14 $\times$ 27 & 1 \\
${\rm [Ne~III]}$ $^{3}P_{1}\rightarrow {^{3}P_{2}}$      & 15.555 & 160.
$\pm$ 32.  & 14 $\times$ 27 & 1\\
${\rm [Fe~II]}$ $^{a4}F_{7/2}\rightarrow {^{a4}F_{9/2}}$ & 17.936 & $<$
10.  & 14 $\times$ 27 & 1\\
${\rm [S~III]}$  $^{3}P_{2}\rightarrow {^{3}P_{1}}$      & 18.713 & 40.   &
14 $\times$ 27 &  1\\
${\rm [Ne~V]}$ $^{3}P_{1}\rightarrow {^{3}P_{0}}$        & 24.317 &
70.  $\pm$ 7.  & 14 $\times$ 27 &  1\\
${\rm [O~IV]}$ $^{2}P_{3/2}\rightarrow {^{2}P_{1/2}}$    & 25.890 &
190.  $\pm$ 20.  & 14 $\times$ 27 &  1\\
${\rm [Fe~II]}$ $^{a6}D_{7/2}\rightarrow {^{a6}D_{9/2}}$ & 25.988 &  8.  &
14 $\times$ 27 & 1\\
${\rm [S~III]}$ $^{3}P_{1}\rightarrow {^{3}P_{0}}$       & 33.481 & 55.  &
20 $\times$ 33 & 1\\
${\rm [Si~II]}$ $^{2}P_{3/2}\rightarrow {^{2}P_{1/2}}$   & 34.814 &
91.     & 20 $\times$ 33 & 1\\
${\rm [Ne~III]}$ $^{3}P_{0}\rightarrow {^{3}P_{1}}$      & 36.013 &
18.     & 20 $\times$ 33 & 1\\
${\rm [O~III]}$ $^{3}P_{2}\rightarrow {^{3}P_{1}}$ & 51.814  & 114. $\pm$
3.   & 80 &  2\\
${\rm [N~III]}$ $^{2}P_{3/2}\rightarrow {^{2}P_{1/2}}$ & 57.317  & 51.4
$\pm$ 2.5  & 80 & 2\\
${\rm [O~I]}$ $^{3}P_{1}\rightarrow {^{3}P_{2}}$ & 63.184  & 156. $\pm$
1.   & 80 & 2\\
${\rm [O~III]}$ $^{3}P_{1}\rightarrow {^{3}P_{0}}$ & 88.356  & 111. $\pm$
1.  & 80 & 2\\
${\rm [N~II]}$ $^{3}P_{2}\rightarrow {^{3}P_{1}}$ & 121.897 & 30.5 $\pm$
1.1 & 80 & 2\\
${\rm [O~I] }$ $^{3}P_{0}\rightarrow {^{3}P_{1}}$ & 145.525 & 11.9 $\pm$
0.4 & 80 & 2\\
${\rm [C~II]}$ $^{2}P_{3/2}\rightarrow {^2P_{1/2}}$ & 157.741 & 216. $\pm$
1.    & 80 & 2\\
\tableline 
OH $^2\Pi _{1/2}$5/2-$^2\Pi _{3/2}$3/2 & 34.60/34.63 & $<$ 3.    & 20
$\times$ 33 & 2\\
OH $^2\Pi _{1/2}$1/2-$^2\Pi _{1/2}$3/2 & 79.11/79.18 & 14.4  $\pm$ 1.5    &
80 & 2\\
OH $^2\Pi _{3/2}$5/2-$^2\Pi _{3/2}$3/2 & 119.23/119.44  & 11.9 $\pm$
1.2   & 80 & 2\\
OH $^2\Pi _{1/2}$3/2-$^2\Pi _{1/2}$1/2 & 163.12/163.40  &  7.42 $\pm$  0.65
& 80 & 2\\
\enddata
\tablecomments{(1): from \citet{lu00} and, where errors are
available, \citet{ale00}; (2): this work}
\end{deluxetable}

\begin{deluxetable}{lccccccc}
\tablecolumns{8} \tablewidth{0pc} \tabletypesize{\scriptsize}
\tablecaption{Comparison of observed line fluxes with AGN model
predictions \label{tbl-2}} \tablehead{ \colhead{Line
id.$\lambda$} &
\multicolumn{7}{c}{Flux (${\rm 10^{-13}~erg~s^{-1}~cm^{-2}}$)}\\
\colhead{($\mu$m )} &
\colhead{Observed/D\tablenotemark{1}~/D\tablenotemark{2}}
  & \multicolumn{2}{c}{AGN A model\tablenotemark{3}} &
\multicolumn{2}{c}{AGN B model\tablenotemark{4}} & \multicolumn{2}{c}{AGN
C model\tablenotemark{5}}\\
  &   &\colhead{Comp. 1}& \colhead{Comp. 2} & \colhead{Comp. 1} &
\colhead{Comp. 2}  & \colhead{Comp. 1}& \colhead{Comp.  2} }
\startdata ${\rm O~VI}$ ${\lambda}$ .1032+.1037 & 37.4/4334./402.
&32.6+19.8   &2.28+1.94    &8.26+5.36 &0.31+0.27 &56.7+31.2
&3.41+2.10  \\
${\rm (Ly \alpha)_{n}}$ $\lambda$ .1215   & 101.8/3562./602    &
179     &  619    &149.     &500     &  239.
&   638. \\
${\rm N~IV]}$ $\lambda$ .1487             &  5.1/103./22.9     &
25.2     &  6.64     &36.3
&6.28       &  59.6   &   47.5  \\
${\rm (CIV)_{n}}$ $\lambda$ .1549         & 39.7/790./177.     &
142.      &  77.6     &110.9      &56.5      &  286.
&    329. \\
${\rm HeII}$ $\lambda$ .1640              & 21.4/426./95.5     &
112.      &  85.9     &108      &81.8      &   346.
&   345. \\
${\rm [Ne~V]}$ $\lambda$ .3426            & 15.7/95./38.7      &
97.7      &  23.3     &68.2      &7.36       &   271.
&    115 \\
${\rm [Ne~III]}$ $\lambda$ .3869+.3968    & 19.2/97./43.2
&37.1+11.2 &41.5+12.5   &102.8+31.0 & 102.+30.7 & 81.6+24.6
&143.+43.0 \\
${\rm HeII}$ $\lambda$ .4686              & 6.1/27.6/13.       &
15.0      &   12.3     &14.7      &11.8      &   46.5
&   48.4  \\
${\rm [O~III]}$ $\lambda$ .4959+.5007     & 256./964./496
&86.9+262
&99.5+294   & 197.+593 & 185.+557. & 148.+446. & 241+726\\
${\rm [Si~VI ]}$ $\lambda$ 1.96           & 8.0/9.2/8.6        &
11.2      &   1.53     &10.8       &1.86       &   23.3  &    3.74  \\%
${\rm [Si~VII ]}$ $\lambda$ 2.48          & 8.3                &
6.41      &   0.1     &9.12      &0.18       &   11.0
&    0.22  \\
${\rm [Si~IX ]}$ $\lambda$ 2.584          & 3.0                &
0.49      & ---    &0.84       & ---   &    0.32
&    ---  \\
${\rm [Mg~VIII]}$ $\lambda$ 3.028         & 11.                &
2.97      & ---    &4.80      & ---   &    4.29
&    ---  \\
${\rm [Si~IX ]}$  $\lambda$ 3.936         & 5.4                &
0.89      & ---    &1.56       & ---   &    0.56
&    ---  \\
${\rm [Mg~IV]}$   $\lambda$ 4.487         & 7.6                &
3.85      &   7.98     &4.29       &6.92       &   9.80
&    19.5  \\
${\rm [Ar~VI]}$   $\lambda$ 4.529         & 15.                &
10.3      &   4.0     &13.4      &3.10       &   26.2
&    20.1  \\
${\rm [Mg~VII]}$  $\lambda$ 5.503         & 13.                &
9.68      &   0.13    &11.4      &.075       &   20.0
&    0.45  \\
${\rm [Mg~V]}$    $\lambda$ 5.610         & 18.                &
10.7      &   8.63    &7.85       &3.84       &   34.6
&   29.5  \\
${\rm [Ar~II]}$   $\lambda$ 6.985         & 13.                &
4.56      &   7.53    &0.93       &1.36       &    0.87
&    1.63  \\
${\rm [Na~III]}$  $\lambda$ 7.318         & 5.8                &
0.56      &   0.59    &1.27       &1.51       &    0.67
&    1.21  \\
${\rm [Ne~VI]}$   $\lambda$ 7.652         & 110.               &
153.7      &   11.9     &105.4      &3.2      &   393.
&    62.4  \\
${\rm [Fe~VII]}$  $\lambda$ 7.815         & 3.0                &
1.83      &   1.63    &1.65       &0.71       &    6.34
&    6.30  \\
${\rm [Ar~V]}$ $\lambda$ 7.902            & $<$ 12.            &
2.14      &   1.83   &2.62       &1.92       &    3.72  &
   6.52  \\
${\rm [Na~VI]}$ $\lambda$ 8.611           & $<$ 16.            &
1.20      &   0.16    &0.95       &---       &    2.95  &
   0.65  \\
${\rm [Ar~III]+[Mg~VII]}$ $\lambda$ 8.991 & 25.                &
4.63+11.7  &7.34+0.17   & 5.97+14.1 & 12.4+0.10   & 4.83+23.8
& 10.7+0.58 \\
${\rm [Fe~VII]}$ $\lambda$ 9.527          & 4.0                &
7.50      &    7.0    &6.75       &3.08       &   25.8
&   26.6  \\
${\rm [S~IV]}$ $\lambda$ 10.510           & 58.                &
38.4      &   60.5    &85.3      &126.      &    75.8
&   187.  \\
${\rm [Ne~II]}$ $\lambda$ 12.813          & 70.                &
5.86      &  20.8     &2.20       &4.44       &    1.62
&    3.26  \\
${\rm [Ar~V]}$ $\lambda$ 13.102           & $<$ 16.            &
2.23      &   2.73    &2.75       &2.88      &    3.93  &
9.57  \\
${\rm [Ne~V]}$ $\lambda$ 14.322           & 97.                &
91.4      &  66.8     &83.2      &28.7      &   270.
&  273.  \\
${\rm [Ne~III]}$ $\lambda$ 15.555         & 160.               &
44.4      &  52.0     &76.     &110.      &   42.0
&   86.9  \\
${\rm [S~III]}$  $\lambda$ 18.713         & 40.                &
25.8      &  56.0     &21.7      &75.7      &   18.4
&   64.5  \\
${\rm [Ne~V]}$ $\lambda$ 24.317           & 70.                &
40.8      &  52.6     &35.8      &22.1      &   116.
&   215.  \\
${\rm [O~IV]}$ $\lambda$ 25.890           & 190.               &
24.1      &  71.9     &28.3      &69.7      &   56.4
&   240.  \\
${\rm [S~III]}$ $\lambda$ 33.481          & 55.                &
8.00      &  29.8     &5.57       &41.1      &    4.84
&   38.4  \\
${\rm [Si~II]}$ $\lambda$ 34.814          & 91.                &
12.9      &  22.6     &6.01       &24.2      &    5.28
&   29.4  \\
${\rm [Ne~III]}$ $\lambda$ 36.013         & 18.                &
3.32      &   4.4     &5.74      &9.40      &   3.20
&    7.46  \\
${\rm [O~III]}$ $\lambda$ 51.814          & 110.               &
9.04      &  33.9     &16.3      &63.2      &   9.06
&   48.5  \\
${\rm [N~III]}$ $\lambda$ 57.317          & 51.                &
2.23      &   11.4     &2.83       &18.9      &    1.88
&   16.1  \\
${\rm [O~I]}$ $\lambda$ 63.184            & 156.               &
3.84      &   1.8     &1.57       &2.26      &    1.24
&    3.10  \\
${\rm [O~III]}$ $\lambda$ 88.356          & 110.               &
1.22      &   8.2     &2.25      &15.5      &    1.23
&   12.6  \\
${\rm [N~II]}$ $\lambda$ 121.897          & 30.                &
0.28      &   0.87    &.07       & 0.40       &    .06
&    0.52  \\
${\rm [O~I] }$ $\lambda$ 145.525          & 12.                &
0.24      &   0.14    &.09       &0.18       &    .07
&    0.24  \\
${\rm [C~II]}$ $\lambda$ 157.741          & 220.               &
0.50      &   1.58     & 0.18       &0.97       &    .14
&    1.09  \\
\enddata
\tablenotetext{1}{Dereddened line flux, assuming E$_{B-V}=0.4$}
\tablenotetext{2}{Dereddened line flux, assuming E$_{B-V}=0.2$}
\tablenotetext{3}{AGN A parameters: component 1: Log~U=-1.,
Log~n=4, internal radius $\simeq$ 21 pc, external radius $\simeq$
109 pc, ionizing spectrum from \citet{ale00}: component 2:
Log~U=-2., Log~n=3.3, internal radius $\simeq$ 153 pc, external
radius $\simeq$ 362 pc, ionizing spectrum from \citet{ale00}.}
\tablenotetext{4}{AGN B parameters: same as AGN A models, but
with the ionizing spectrum from \citet{pi94}}
\tablenotetext{5}{AGN C parameters: same as AGN A models, but
with the ionizing spectrum that includes a big blue bump (see
text)}

\end{deluxetable}

\begin{deluxetable}{lccccccc}
\tabletypesize{\scriptsize} \tablecaption{Comparison of observed
line fluxes with the Ring Starburst model predictions
\label{tbl-3}} \tablewidth{0pt}
\tablehead{ \colhead{Line id.$\lambda$} &
\multicolumn{7}{c}{Flux (${\rm 10^{-13}~erg~s^{-1}~cm^{-2}}$)}\\
\colhead{($\mu$m )} & \colhead{Observed} &\colhead{SBR
A\tablenotemark{1}} &\colhead{SBR B\tablenotemark{2}}
&\colhead{SBR C\tablenotemark{3}} &\colhead{SBR
D\tablenotemark{4}} &\colhead{SBR E\tablenotemark{5}} &
\colhead{SBR F\tablenotemark{6}}} \startdata
${\rm O~VI}$ ${\lambda}$ .1032+.1037      &37.4/4334./402.    &  ---   &   ---  &   ---  &   ---     & ---    & ---     \\
${\rm (Ly \alpha)_{n}}$ $\lambda$ .1215   &101.8/3562./602    &  1330.  &  865.  &  997.  &  875.    &  1010. & 917.   \\
${\rm N~IV]}$ $\lambda$ .1487             &  5.1/103./22.9    & 71.9  &  0.20  &  85.8  &  .020 &  93.7  & 0.03    \\
${\rm (CIV)_{n}}$ $\lambda$ .1549         &39.7/790./177.     & 795.  &  0.70  &  858.  &   0.80     &  924. & 0.87    \\
${\rm HeII}$ $\lambda$ .1640              &21.4/426./95.5     & 693.  &  50.2  &  696.  & 50.5       &  696. & 50.8    \\
${\rm [Ne~V]}$ $\lambda$ .3426            &15.7/95./38.7      &  58.1  &  ---   &  69.6  &   ---     &  73.9  & ---     \\
${\rm [Ne~III]}$ $\lambda$ .3869+.3968    &19.2/97./43.2      &314.+94.7 & 31.1+9.4 & 309.+93.1 &  31.5+9.5  & 316.+95.7 & 32.+9.7 \\
${\rm HeII}$ $\lambda$ .4686              &6.1/27.6/13.       &  97.3  &  7.4  & 97.7  &  7.4     &  97.3  & 7.4    \\
${\rm [O~III]}$ $\lambda$ .4959+.5007     & 256./964./496     &1330.+3990. & 25.1+75.6 & 1383+4158.& 26.0+78.2& 1412.+4257. & 26.8+80.5 \\
${\rm [Si~VI ]}$ $\lambda$ 1.96           &8.0/9.2/8.6        &  ---   &  ---   &  ---   &   ---     &  ---   &  ---    \\
${\rm [Si~VII ]}$ $\lambda$ 2.48          &8.3                &  ---   &  ---   &  ---   &   ---     &  ---   &  ---    \\
${\rm [Si~IX ]}$ $\lambda$ 2.584          &3.0                &  ---   &  ---   &  ---   &   ---     &  ---   &  ---    \\
${\rm [Mg~VIII]}$ $\lambda$ 3.028         &11.                &  ---   &  ---   &  ---   &   ---     &  ---   &  ---    \\
${\rm [Si~IX ]}$  $\lambda$ 3.936         &5.4                &  ---   &  ---   &  ---   &   ---     &  ---   &  ---    \\
${\rm [Mg~IV]}$   $\lambda$ 4.487         &7.6                &  3.30   &   ---  & 3.50   &   ---     &  3.53   &  ---    \\
${\rm [Ar~VI]}$   $\lambda$ 4.529         &15.                &  6.20   &   ---  &  7.72   &   ---     &  8.02  &  ---    \\
${\rm [Mg~VII]}$  $\lambda$ 5.503         &13.                &  ---   &   ---  &  ---   &   ---     &  ---   &  ---    \\
${\rm [Mg~V]}$    $\lambda$ 5.610         &18.                &  3.10  &   ---  &  3.63   &   ---     &  3.73   &  ---    \\
${\rm [Ar~II]}$   $\lambda$ 6.985         &13.                &  5.90   &  3.70   &  4.29   &   3.70     &  4.12  &  3.73  \\
${\rm [Na~III]}$  $\lambda$ 7.318         &5.8                &  0.60   &  0.10  &  0.54   &   0.11     &  0.54   &  0.11    \\
${\rm [Ne~VI]}$   $\lambda$ 7.652         &110.               &  16.9   &  ---   &  21.1   &   ---     &  21.8   &   ---   \\
${\rm [Fe~VII]}$  $\lambda$ 7.815         &3.0                &  0.40  &  ---   &  0.48   &   ---     &  0.50   &   ---   \\
${\rm [Ar~V]}$ $\lambda$ 7.902            & $<$12.            &  7.50   &  ---   & 8.84   &   ---     &  9.11  &   ---   \\
${\rm [Na~VI]}$ $\lambda$ 8.611           & $<$16.            &  ---   &  ---   &   ---  &   ---     &  ---   &   ---   \\
${\rm [Ar~III]+[Mg~VII]}$ $\lambda$ 8.991 &25.                &  46.5   &  7.10   & 43.2   &   7.13   &  42.2  &  7.13  \\
${\rm [Fe~VII]}$ $\lambda$ 9.527          &4.0                &  1.70  &  ---   &  2.04   &   ---     &  2.11   &   ---   \\
${\rm [S~IV]}$ $\lambda$ 10.510           &58.                &  312.  &  0.80   &  346.  &   0.84     &  333.  &  0.81    \\
${\rm [Ne~II]}$ $\lambda$ 12.813          &70.                &  7.80   & 5.50   &  5.44  &   5.38     &  5.38  &  5.61   \\
${\rm [Ar~V]}$ $\lambda$ 13.102           & $<$16.            &  21.9   &  ---   &  25.5 &   ---     &  24.9  &   ---   \\
${\rm [Ne~V]}$ $\lambda$ 14.322           & 97.                &271.   &  ---   &  315.  &   ---     &  314.  &   ---   \\
${\rm [Ne~III]}$ $\lambda$ 15.555         & 160.               &295.   &  50.5  &  276.  &   50.2    &  270.  &  49.8   \\
${\rm [S~III]}$  $\lambda$ 18.713         & 40.                &180.   &  19.2  &   178.  &   20.3    &  203.  &  23.7   \\
${\rm [Ne~V]}$ $\lambda$ 24.317           & 70.                &303.   &  ---   &  346.  &   ---     &  294.  &  ---    \\
${\rm [OIV]}$ $\lambda$ 25.890           & 190.                &1660.   &   1.90  & 1762.  &    2.00    & 1419.   & 1.62   \\
${\rm [S~III]}$ $\lambda$ 33.481          & 55.                &785.   &   87.5  &  673.  &   81.2    &  350.  &  46.2  \\
${\rm [Si~II]}$ $\lambda$ 34.814          & 91.                &126.   &   54.8  &  89.4  &   49.5    &  41.9  &  29.4  \\
${\rm [Ne~III]}$ $\lambda$ 36.013         & 18.                &52.8   &   8.90 &   49.2  &    8.91    &   47.5  &   8.74  \\
${\rm [O~III]}$ $\lambda$ 51.814          & 110.               &2980.   &   81.5  & 3217  &   87.5    & 2524.  &  71.9  \\
${\rm [N~III]}$ $\lambda$ 57.317          & 51.                &1110.   &   40.0  & 921  &   35.0    &  372.9  &  14.7  \\
${\rm [O~I]}$ $\lambda$ 63.184            &156.               & 165.   &   130.  &   113. &   127.    &  116.  &  124.  \\
${\rm [O~III]}$ $\lambda$ 88.356          & 110.               &4980.  &  138.  & 3729. &  109.    & 977.  &   29.0  \\
${\rm [N~II]}$ $\lambda$ 121.897          &30.                & 80.0  &   37.3 &  47.8 &   29.7    &   12.3 &   8.78  \\
${\rm [O~I] }$ $\lambda$ 145.525          & 12.                &16.2    &   12.9  &   11.0  &   12.5     &  10.1  &   11.3  \\
${\rm [C~II]}$ $\lambda$ 157.741          & 220.               &739.   &   465. &  181.  &   172.    & 67.6  & 46.5  \\
\enddata
\tablecomments{*: this line was used for normalization }

\tablenotetext{1}{SBR A parameters: Log~U=-2.5, Log~n=1.0,
ionizing spectrum from Starburst99 with instantaneous
star-formation law, M = 10$^6 M_{\sun}$, IMF: =2.35 M$_{up} = 100
M_{\sun}$, M$_{low} = 1 M_{\sun}$, nebular emission included,
Z=0.020, age of 5 Myr. The integration was stopped at a
temperature of 50K, the adopted abundances are those relative to
HII regions and grain emission is included. The adopted
number of clouds is 33000.}

\tablenotetext{2}{SBR B parameters: Log~U=-3.5, Log~n=1.0, all
other parameters as for SBR A. The number of clouds
adopted is 33000}

\tablenotetext{3}{SBR C parameters: Log~U=-2.5, Log~n=2.0, all
other parameters as for SBR A. The number of clouds
adopted is 3300}

\tablenotetext{4}{SBR D parameters: Log~U=-3.5, Log~n=2.0, all
other parameters as for SBR A. The number of clouds
adopted is 3300}

\tablenotetext{5}{SBR E parameters: Log~U=-2.5, Log~n=3.0, all
other parameters as for SBR A. The number of clouds
adopted is 330}

\tablenotetext{6}{SBR F parameters: Log~U=-3.5, Log~n=3.0, all
other parameters as for SBR A. The number of clouds
adopted is 330}
\end{deluxetable}

\begin{deluxetable}{lcccc}
\tabletypesize{\scriptsize}
\tablecaption{Comparison of observed line fluxes with composite
model predictions \label{tbl-4}} \tablewidth{0pt}
\tablehead{ \colhead{Line id.$\lambda$} &
\multicolumn{4}{c}{Flux (${\rm 10^{-13}~erg~s^{-1}~cm^{-2}}$)}\\
\colhead{($\mu$m )} &
\colhead{Observed/D\tablenotemark{1}~/D\tablenotemark{2}}
&\colhead{CM1\tablenotemark{3}} &\colhead{CM2\tablenotemark{4}}
&\colhead{CM3\tablenotemark{5}}} \startdata
${\rm O~VI}$ $\lambda$ .1032+.1037   & 37.4/4334./402.        &  56.6    &   14.2  &   93.4 \\
${\rm (Ly \alpha)_{n}}$ $\lambda$ .1215   & 101.8/3562./602   & 1673.    &   1524. &  1752. \\
${\rm N~IV]}$ $\lambda$ .1487          &5.1/103./22.9         &  31.8    &   42.6  &  107.  \\
${\rm (CIV)_{n}}$ $\lambda$ .1549      &39.7/790./177.        &  220.    &   168.  &  616. \\
${\rm HeII}$ $\lambda$ .1640           &21.4/426./95.5        &  248.    &   240.  &   741. \\
${\rm [Ne~V]}$ $\lambda$ .3426         &15.7/95./38.7         &  121.    &   75.6  &   386. \\
${\rm [Ne~III]}$ $\lambda$ .3869+.3968 & 19.2/97./43.2        &  143.    &   307.  &   333. \\
${\rm HeII}$ $\lambda$ .4686           &6.1/27.6/13.          &  34.7    &   33.9  & 102.3  \\
${\rm [O~III]}$ $\lambda$ .4959+.5007  & 256./964./496        &  847.    &   1636.  & 1665.  \\
${\rm [Si~VI ]}$ $\lambda$ 1.96        &8.0/9.2/8.6           &  12.7    &   12.7  &  27.0  \\
${\rm [Si~VII ]}$ $\lambda$ 2.48       &8.3                   &   6.5    &   9.3  &  11.2  \\
${\rm [Si~IX ]}$ $\lambda$ 2.584       &3.0                   &   0.5    &    0.8  &   0.3  \\
${\rm [Mg~VIII]}$ $\lambda$ 3.028      &11.                   &   3.0    &   4.8  &   4.3  \\
${\rm [Si~IX ]}$  $\lambda$ 3.936      &5.4                   &   0.9    &    1.6  &   0.6  \\
${\rm [Mg~IV]}$   $\lambda$ 4.487      &7.6                   &   11.8    &   11.2  &  29.3  \\
${\rm [Ar~VI]}$   $\lambda$ 4.529      &15.                   &   14.3    &   16.5  &  46.3  \\
${\rm [Mg~VII]}$  $\lambda$ 5.503      &13.                   &   9.8    &  11.5   &   20.5 \\
${\rm [Mg~V]}$    $\lambda$ 5.610      &18.                   &   19.3    &   11.7   & 64.1 \\
${\rm [Ar~II]}$   $\lambda$ 6.985      &13.                   &   15.8    &   6.0   &   6.2  \\
${\rm [Na~III]}$  $\lambda$ 7.318      &5.8                   &   1.3    &   2.8   &   2.0  \\
${\rm [Ne~VI]}$   $\lambda$ 7.652      &110.                  &   166.    &  109.   &   455. \\
${\rm [Fe~VII]}$  $\lambda$ 7.815      &3.0                   &   3.5    &   2.4   &    12.6 \\
${\rm [Ar~V]}$ $\lambda$ 7.902         & $<$12.               &   4.0    &   4.5   &   10.2 \\
${\rm [Na~VI]}$ $\lambda$ 8.611        & $<$16.               &   1.4    &   0.9   &    3.6 \\
${\rm [Ar~III]+[Mg~VII]}$ $\lambda$ 8.991 &25.                &   31.    &  39.7   &   47.0 \\
${\rm [Fe~VII]}$ $\lambda$ 9.527       &4.0                   &   14.5    &  9.8   &   52.4 \\
${\rm [S~IV]}$ $\lambda$ 10.510        &58.                   &   99.7    &  212.   &   263. \\
${\rm [Ne~II]}$ $\lambda$ 12.813       &70.                   &   32.0    &  12.0   &   10.3 \\
${\rm [Ar~V]}$ $\lambda$ 13.102    & $<$16.                   &   5.0    &   5.6   &    13.5 \\
${\rm [Ne~V]}$ $\lambda$ 14.322        &97.                   &   158.   &  112.   &   543. \\
${\rm [Ne~III]}$ $\lambda$ 15.555      & 160.                  &   147.   & 236.   &   179. \\
${\rm [S~III]}$  $\lambda$ 18.713      & 40.                   &   102.   & 118.   &   103. \\
${\rm [Ne~V]}$ $\lambda$ 24.317        &70.                   &   93.4    &  57.9   &   331. \\
${\rm [O~IV]}$ $\lambda$ 25.890        &190.                  &   98.    &  100.   &   298. \\
${\rm [S~III]}$ $\lambda$ 33.481       & 55.                   &   119.  & 128.   &   124. \\
${\rm [Si~II]}$ $\lambda$ 34.814       &91.                   &   85.   &  79.7   &   84.2 \\
${\rm [Ne~III]}$ $\lambda$ 36.013      &18.                   &   16.6   &  24.0   &   19.6  \\
${\rm [O~III]}$ $\lambda$ 51.814       & 110.                  &   130.   & 167.   &   145. \\
${\rm [N~III]}$ $\lambda$ 57.317       &51.                   &   48.6    &  56.7   &   53.  \\
${\rm [O~I]}$ $\lambda$ 63.184         & 156.                  &   133.   & 131.   &   131. \\
${\rm [O~III]}$ $\lambda$ 88.356       & 110.                  &   118.   & 127.   &   123. \\
${\rm [N~II]}$ $\lambda$ 121.897       &30.                   &   30.8    &  30.   &   30.3 \\
${\rm [O~I] }$ $\lambda$ 145.525       &12.                   &   12.9   &  12.8   &   12.8 \\
${\rm [C~II]}$ $\lambda$ 157.741       & 220.                  &   174.   & 173.   &   173.  \\
\enddata
{\tablenotetext{1}{Dereddened line flux, assuming
E$_{B-V}=0.4$} \tablenotetext{2}{Dereddened line flux, assuming
E$_{B-V}=0.2$} \tablenotetext{3}{CM1 = AGN A + SBR D}
\tablenotetext{4}{CM2 = AGN B + SBR D} \tablenotetext{5}{CM3 =
AGN C + SBR D}}
\end{deluxetable}

\begin{deluxetable}{lccc}
\tabletypesize{\scriptsize} \tablecaption{Comparison of observed
line fluxes with model predictions for the nuclear region
\label{tbl-5}} \tablewidth{0pt} \tablehead{ \colhead{Line
id.$\lambda$} &
\multicolumn{2}{c}{Flux ($10^{-12}~erg~s^{-1}~cm^{-2}$)} & \colhead{Notes}\\
\colhead{}& \colhead{Observed}   & \colhead{Modeled} & \colhead{}}
\startdata
34$\mu$m   &  $<$ 0.1 & -0.43 & (absorption)\\
53$\mu$m   &  $<$ 1.2 &  -0.66 & (absorption)\\
79$\mu$m   &  1.1    &  1.13 & \\
84$\mu$m   &  $<$ 1.2 &  0.06  &  \\
98$\mu$m   &  $<$ 1.2 &  0.17  &  \\
119$\mu$m  &  1.3    & 1.60  &  \\
163$\mu$m  &  0.38    & 0.35  &  \\
\enddata
\end{deluxetable}

\end{document}